\documentclass[prb, twocolumn, showpacs, amsmath, amssymb, superscriptaddress, bibnotes]{revtex4-2}

\usepackage{graphicx}
\usepackage{dcolumn}
\usepackage{bm}
\usepackage{xcolor}
\usepackage{bbold}
\usepackage{float}
\usepackage{soul}
\usepackage{enumitem}
\usepackage[protrusion=true,expansion=true,final]{microtype}
\usepackage{comment}
\usepackage[normalem]{ulem}

\usepackage{tikz}

\usetikzlibrary{calc}
\usetikzlibrary{decorations.pathmorphing}
\usetikzlibrary{decorations.pathreplacing}
\usetikzlibrary{decorations.markings}
\usetikzlibrary{shapes.misc}
\usetikzlibrary{positioning}
\usetikzlibrary{snakes}
\usetikzlibrary{arrows}
\usetikzlibrary{fadings}
\usetikzlibrary{shapes.callouts,tikzmark}
\tikzfading[name=fade out,
inner color=transparent!0,
outer color=transparent!100]
\tikzset{middlearrow/.style={
                decoration={markings,
                            mark= at position 0.5 with {\arrow{#1}} ,
                                    },
                                            postaction={decorate}
                                                }
                                                }
\tikzset{endarrow/.style={
                decoration={markings,
                            mark= at position 0.9 with {\arrow{#1}} ,
                                    },
                                            postaction={decorate}
                                                }
                                                }
\tikzset{startarrow/.style={
                decoration={markings,
                            mark= at position 0.1 with {\arrow{#1}} ,
                                    },
                                            postaction={decorate}
                                                }
                                                }

\usepackage{hyperref}
\hypersetup{colorlinks = true, urlcolor = blue, linkcolor = blue, citecolor = blue}

\newcommand{\ph}{\text{ph}}
\newcommand{\pp}{\text{pp}}
\newcommand{\cph}{\overline{\text{ph}}}

\newcommand{\charge}{\text{ch}}
\newcommand{\spin}{\text{sp}}
\newcommand{\singlet}{\text{s}}
\newcommand{\triplet}{\text{t}}

\newcommand{\pol}{P}
\newcommand{\DCAsq}{\text{DCA}_4}
\newcommand{\DCAst}{\text{DCA}_{4^*}}

\newcommand{\FK}[1]{{\color{red}#1}}

\begin{document}

\title{Embedded multi-boson exchange: A step beyond quantum cluster theories}

\author{Dominik Kiese}
    \affiliation{Center for Computational Quantum Physics, Flatiron Institute, 162 5th Avenue, New York, NY 10010, USA}

\author{Nils Wentzell}
    \affiliation{Center for Computational Quantum Physics, Flatiron Institute, 162 5th Avenue, New York, NY 10010, USA}

\author{Igor Krivenko}
    \affiliation{Institut f\"ur Theoretische Physik, Universit\"at Hamburg, Notkestraße 9 , 22607 Hamburg, Germany}

\author{Olivier Parcollet}
    \affiliation{Center for Computational Quantum Physics, Flatiron Institute, 162 5th Avenue, New York, NY 10010, USA}
    \affiliation{Universit\'e Paris-Saclay, CNRS, CEA, Institut de physique th\'eorique, 91191, Gif-sur-Yvette, France}

\author{Karsten Held}
    \affiliation{Institute for Solid State Physics, TU Wien, 1040 Vienna, Austria}

\author{Friedrich Krien}
    \affiliation{Institute for Solid State Physics, TU Wien, 1040 Vienna, Austria}

\date{\today}

\begin{abstract}
    We introduce a diagrammatic multi-scale approach to the Hubbard model based on the interaction-irreducible (multi-boson) vertex
    of a small cluster embedded in a self-consistent medium. The vertex captures short-ranged correlations up to the length scale of the cluster,
    while long-ranged correlations are recovered from a set of diagrammatic equations for the Hedin three-leg vertex.
    By virtue of the crossing symmetry, the Fierz decoupling ambiguity of the Hubbard interaction is resolved exactly.
    Our benchmarks for the half-filled Hubbard model on the square lattice are in very good agreement with numerically exact diagrammatic Monte Carlo simulations. 
\end{abstract}

\maketitle

\section{Introduction}

Quantum embedding methods, such as dynamical mean-field theory
(DMFT)~\cite{Georges_1996, Kotliar_2006, Pavarini_2011}, have had a profound
impact on our understanding of strongly correlated systems.  DMFT constructs an
approximation of the quantum many-body problem based on a local approximation
of the self-energy using an auxiliary Anderson impurity model embedded in a
self-consistent, non-interacting bath.  While it is very successful in
describing local quantum fluctuations, DMFT suffers from several limitations,
in particular the lack of momentum dependence of the self-energy or 
feedback of the long range fluctuations onto the electronic degrees of freedom.  In
order to overcome those limitations, two main directions have been explored in
recent years.

First, cluster approaches~\cite{Maier_2005} enlarge the impurity problem to $N_c>1$ sites.
They describe short range quantum fluctuations, reintroduce some coarse grained momentum dependency 
of the self-energy and allow one to compute non-local order parameters, e.g. d-wave superconducting order. 
They also provide some systematic control, as $N_c=1$ corresponds to DMFT while $N_c=\infty$ is the exact solution.
The calculation of large clusters is however quite limited (except at high temperatures)
due to the limitations of the quantum impurity solvers (in particular the fermionic sign problem in quantum Monte Carlo).
As a result, some weak coupling mechanisms involving the effect of 
long-ranged correlations on electronic degrees of freedom are out of reach for these methods.
For example, in the square-lattice Hubbard model at weak coupling $U/t$ (where $U$ is the on-site Coulomb repulsion and $t$ is the hopping amplitude),
spin fluctuations open a pseudogap at low temperature~\cite{Vilk_1997}. A proper cluster
treatment of this effect at $U=2t$ would require at least $1000$ sites~\cite{Schaefer_2021}, which is unrealistic to achieve for cluster DMFT methods.

Second, diagrammatic extensions of DMFT~\cite{Rohringer_2018} constitute another family of quantum embeddings.
Beyond DMFT, they use some non-local diagrams to calculate non-local self-energies.
The simplest such approach is GW+DMFT~\cite{Biermann_2003}.
While DMFT is a local approximation of the self-energy,
many diagrammatic extensions use a local approximation for some vertex, e.g.  
the triply irreducible local expansion (TRILEX,~\cite{Ayral_2016,Ayral_2017}),
the dynamical vertex approximation (D$\Gamma$A,~\cite{Toschi_2007,Katanin2009}), 
ladder approximations~\cite{Toschi_2007,Hafermann_2009} and parquet schemes~\cite{Valli2015,Astretsov_2020,Krien_2020_1,Krien_2021}.
Some of them are also based on dual fermions~\cite{Rubtsov_2008}.
Similarly, non-local correlations beyond DMFT may be taken into account through the functional renormalization group (fRG)~\cite{Taranto_2014}.
Diagrammatic extensions of single-site DMFT allow for a fine momentum resolution~\footnote{except for the parquet approaches that are very memory-intensive},
and the study of the pseudogap  at weak coupling~\cite{Schaefer_2015,vanLoon_2018}
but possibly miss some important short range effects, such as the formation of singlets~\cite{Hafermann_2008}. Furthermore, conservation laws and the Pauli principle cannot be fulfilled at the same time~\cite{Smith_1992,Kugler_2018,Green_2024}.
Fulfilling the latter, thermodynamic inconsistencies are thus unavoidable~\cite{vanLoon_2016,Janis_2017,Krien_2017}. 

Cluster and diagrammatic extensions of DMFT are therefore quite complementary.
They can be combined to  obtain a {\it multi-scale} approach,
using a cluster instead of a single site as a starting point of the diagrammatic extension.
Clusters provide short range correlations and control to the methods, while the diagrammatic extension
accounts for the long range correlations. 
In practice, few cases have been studied up to now, e.g. cluster
TRILEX~\cite{Ayral_2017}, low-order diagrams within the dual fermion
formalism~\cite{Iskakov_2018,vanLoon_2021,Iskakov_2024},
or ladder approximations beyond the dynamical cluster approximation (DCA,~\cite{Slezak_2009,Yang_2011}) and beyond cluster DMFT~\cite{Hafermann_2008}.
This is mainly due to the rapid growth of computational cost with the cluster size $N_c$.
Furthermore, the key question of the convergence speed of various methods with $N_c$ has not been systematically explored yet.

In this paper, we propose and explore a method based on a local (cluster) approximation of the interaction-irreducible vertex $\Lambda^U$ of the so-called
single-boson exchange (SBE) decomposition~\cite{Krien_2019}.
Since this quantity corresponds diagrammatically to the repeated exchange of bosons~\cite{Krien_2020_1,Krien_2021}, 
we call this approach \emph{embedded multi-boson exchange} (eMBEX).
Here \emph{embedded} bears a double meaning: {\it (i)} the cluster on which  $\Lambda^U$ is calculated is embedded in a bath, and
{\it(ii)} $\Lambda^U$ is embedded into the lattice via the SBE equations.
A cluster approximation to $\Lambda^U$ is a promising route,
because there is evidence that this quantity remains short ranged
even in the presence of long-ranged antiferromagnetic correlations~\cite{Krien_2020_1,Krien_2020_2,Krien_2022_2}.
In practice, we choose DCA~\cite{Hettler_1998,Maier_2005} as the cluster method, 
with a cluster size of $N_c=4$.
We show that eMBEX yields very good results in benchmarks against the numerically exact diagrammatic Monte Carlo (DiagMC)~\cite{Simkovic_2019}
for the electronic self-energy. We also discuss the
effect of different patch systems of the Brillouin zone in the DCA cluster.

Compared to a cluster parquet D$\Gamma$A approach, eMBEX has two main advantages.
First, since the SBE decomposition makes use of physical correlation functions, 
it does not experience the divergences observed in the fully irreducible 4-leg vertex~\cite{Schaefer_2013,Gunnarsson_2016,Thunstroem_2018,Chalupa_2018,Chalupa_2021}.
Second, solving the D$\Gamma$A parquet equations of the lattice is a much more formidable task, thus restricting the 
size of the ${\bm{k}}$-mesh considerably \cite{Li2017,Eckhardt_2020}.
The eMBEX equations  require much fewer computational resources~\cite{Krien_2019_2}, as more involved multi-boson diagrams are restricted to the cluster, and thus allow for a finer 
${\bm{k}}$-mesh.
Still, similar to the parquet schemes, approximations to $\Lambda^U$ satisfy the crossing symmetry and thus the Pauli principle.
\begin{figure*}
\begin{center}
\includegraphics[width = 0.8\linewidth]{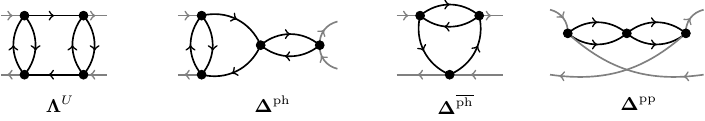}
\end{center}
    \caption{\label{fig:reducibility}\textbf{Example of Feynman diagrams} from the irreducible $(\bm{\Lambda}^U)$ and reducible $(\bm{\Delta})$
    classes of the SBE decomposition. Filled circles denote the bare interaction, arrows the Green's function,
    black (gray) marks internal (external) propagators.
    }
\end{figure*}

The paper is structured as follows:~in Sec.~\ref{sec:SBE_decomposition} we
briefly review the single-boson exchange decomposition of the two-particle
vertex. We introduce the central eMBEX approximation of  $\Lambda^U$ in Sec.~\ref{sec:parametrization}. The numerical
implementation of our method is outlined in Secs.~\ref{sec:SBE_equations} and
\ref{sec:implementation}. Benchmarks for the half-filled Hubbard model on the
square lattice are presented in Sec.~\ref{sec:results}. We conclude in
Sec.~\ref{sec:conclusion}.

\section{SBE decomposition}
\label{sec:SBE_decomposition}

The SBE decomposition introduced in Ref.~\cite{Krien_2019} corresponds to a classification of the vertex diagrams into various
reducible or irreducible classes, similar to the parquet decomposition~\cite{Bickers_2004}.
In the latter case, diagrams are (ir)reducible with respect to cutting pairs of Green's function lines (`$GG$-(ir)reducible').
A $GG$-reducible diagram can be either reducible in the horizontal ($\ph$) or vertical ($\cph$) particle-hole channel,
or in the particle-particle ($\pp$) channel, or it is $GG$-irreducible. 

In the SBE decomposition, on the other hand, (ir)reducibility implies that a diagram can (not)
be split into two parts by removing a bare interaction vertex (`$U$-(ir)reducible'), see also Fig.~\ref{fig:reducibility}, where $U$ denotes the Hubbard interaction.
With the exception of the bare interaction itself, each $U$-reducible diagram belongs to one
and only one of the three $GG$-reducible parquet classes~\cite{Krien_2019,Gievers_2022}.

As a result, the full vertex may be written as a sum of the $U$-reducible and -irreducible diagrams (cf.\ Fig.~\ref{fig:reducibility}):
\begin{eqnarray}
    \bm{F}=\bm{\Lambda}^{U}+\sum_r\bm{\Delta}^r-2\bm{U},\;\;\;r \in \{\ph,\cph,\pp\}.\label{eq:sbe_decomposition}
\end{eqnarray}
Here, $\bm{\Lambda}^{U}$ is the $U$-irreducible residual vertex of the SBE decomposition;
the $U$-reducible diagrams are denoted as $\bm{\Delta}^\ph, \bm{\Delta}^{\cph},$ and $\bm{\Delta}^\pp$.
The last term $2\bm{U}$ is a double-counting correction, taking into account that the bare interaction arises to leading order from each $\bm{\Delta}$.

For simplicity we adopt a matrix notation using bold symbols:
each vertex is initially a tensorial quantity, such as $F^{\ell_1\cdots\ell_4}$ ($\ell_i$ is a fermionic multi-index including frequency, momentum, spin and possibly orbital degrees of freedom),
which we denote as a matrix $(\bm{F})^{ab}$ with $a=(\ell_1,\ell_2)$ and $b=(\ell_3,\ell_4)$.
Further below we focus on the single-band case, but up to this point the notation generalizes to multiple orbitals.

The key observation of Ref.~\cite{Krien_2019} is that the $U$-reducible diagrams may be written
in terms of the Hedin vertices $\bm{\gamma}$,$\overline{\bm{\gamma}}$~\footnote{The vertices $\bm{\gamma}$ and $\overline{\bm{\gamma}}$ only differ by their respective ordering of fermionic and bosonic legs.} and the screened interaction $\bm{W}$:
\begin{eqnarray}
    \bm{\Delta}^r=\overline{\bm{\gamma}}^r\bm{W}^r\bm{\gamma}^r.\label{eq:delta}
\end{eqnarray}
This corresponds to a \emph{single} bosonic line $\bm{W}$
enclosed between two three-leg vertices, hence the name SBE decomposition.
As discussed in the literature~\cite{Ayral_2016,Krien_2019},
$\bm{\gamma}$ and $\bm{W}$ are connected to fermion-boson and boson-boson correlation functions, respectively.

In practice, when $\bm{F}$ and the $\bm{\Delta}$'s are known, the vertex $\bm{\Lambda}^U$ is obtained from Eq.~\eqref{eq:sbe_decomposition}.
Note that $\bm{W}, \bm{\gamma},$ and $\bm{\Lambda}^U$ depend on one, two,
and three frequencies and momenta, respectively, as discussed in the following.

\section{Approximation of $\Lambda^U$}
\label{sec:parametrization}

The SBE decomposition~\eqref{eq:sbe_decomposition} shows that the full vertex $\bm{F}$ receives a feedback
from the screened interaction $\bm{W}$, cf. Eq.~\eqref{eq:delta}.
In a regime of long-range antiferromagnetic correlations $\bm{W}$
is long ranged in the spin sector and therefore also $\bm{F}$ is long ranged.
For $\bm{\Lambda}^U$ the situation is less clear, albeit
previous investigations by some of the authors suggest that an expansion of $\bm{\Lambda}^U$ in terms of form factors converges fast even when $\bm{W}$ is long ranged~\cite{Krien_2020_1,Krien_2020_2,Krien_2022_2}.

\begin{figure}
    \centering
    \includegraphics[width=0.9\linewidth]{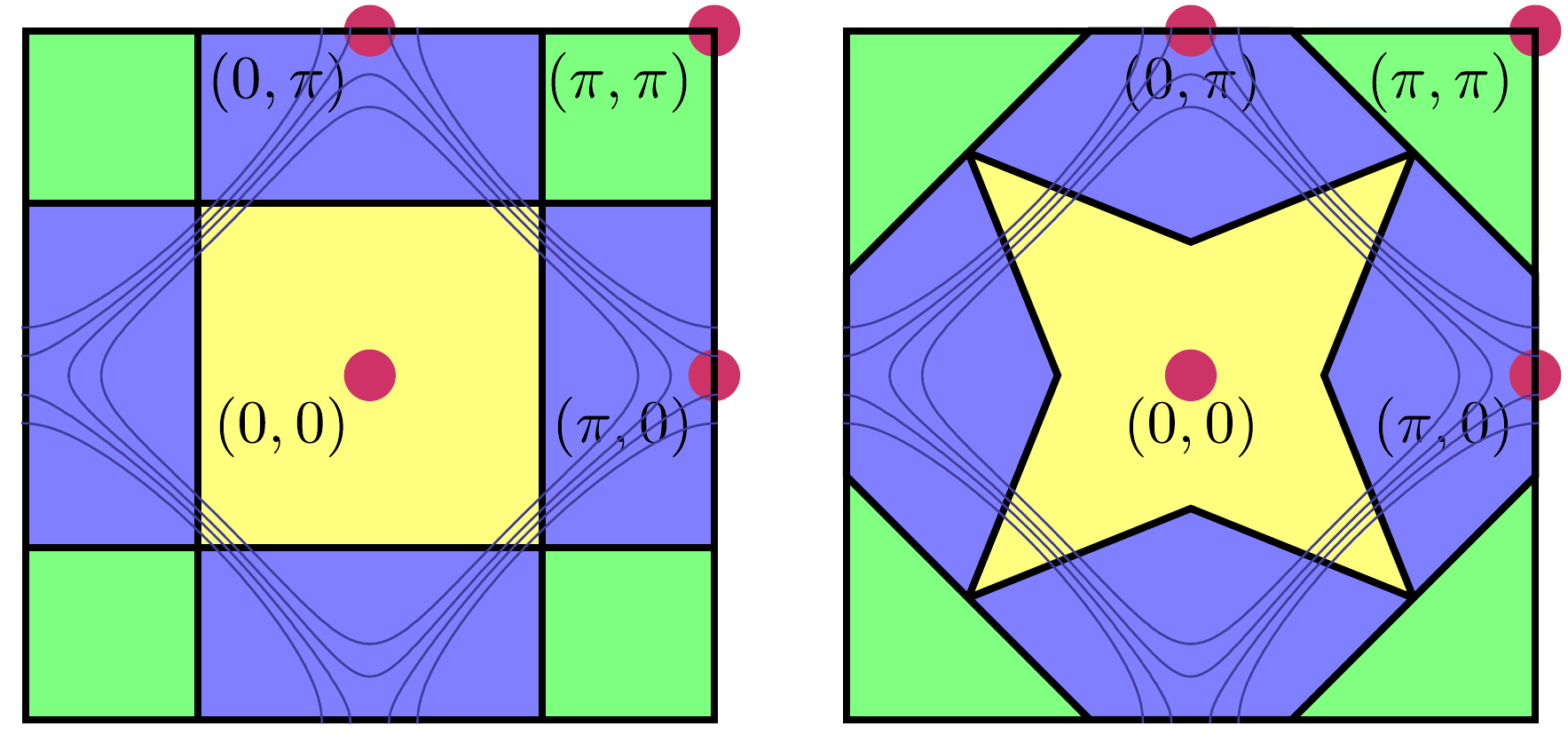}
    \caption{\textbf{DCA patching geometries.} The left-hand side shows the square patching ($\DCAsq$) whereas the right-hand side corresponds to the star patching ($\DCAst$). Figure reproduced from Ref.~\cite{Gull_2010}.}
    \label{fig:patching_geometries}
\end{figure}

This motivates the approximation employed in the present work:~the lattice vertex $\bm{\Lambda}^{U, \text{lat}}$ is approximated by the respective vertex of a self-consistently embedded cluster with $N_c$ sites,
\begin{align}
    \bm{\Lambda}^{U, \text{lat}} \approx \bm{\Lambda}^{U, c} \,.
\end{align}
We consider translationally invariant clusters with $N_c = 1$ and $N_c = 4$, corresponding to single-site DMFT and 4-site dynamical cluster approximation calculations, respectively. In our DCA calculations we consider two different schemes to tile the Brillouin zone into patches: the \emph{square} patching, which we refer to as $\DCAsq$ and the \emph{star} patching (see Fig.~\ref{fig:patching_geometries}) denoted by $\DCAst$ in the following \cite{Gull_2010}.

For $N_c > 1$, a practical problem arises in the parametrization of the lattice vertex $\bm{\Lambda}^{U, \text{lat}}_{\bm{k}_1 \bm{k}_2 \bm{k}_3}$,
because the cluster quantity $\bm{\Lambda}^{U,c}_{\bm{K}_1 \bm{K}_2 \bm{K}_3}$ is defined only on the coarse-grained momenta $\bm{K}$. The most straightforward solution consists in extending the DCA approximation for the self-energy to the vertex, i.e.~$\Lambda^{U, \text{lat}}$ is treated as a constant function of momentum on the cluster patches. Unfortunately, such an approximation spoils the smoothness of lattice quantities computed from $\Lambda^{U, \text{lat}}$ and we therefore do not consider this approach in the following. Another possible solution is the DCA+~\cite{Pavarini_2015} where one-particle~\cite{Staar_2013}
and two-particle quantities~\cite{Staar_2014} on the fine grid are obtained by
formally inverting the convolution defined through the patching function $\Theta$ (cf.~Sec.~\ref{subsec:DCA}). 

Here, we recover $\bm{\Lambda}^{U, \text{lat}}$ as a smooth function of three momenta using the available cluster vertex data. To achieve this, we treat the representation of the irreducible vertex on the lattice as an interpolation problem~\footnote{We require $\bm{\Lambda}^{U, \text{lat}}_{\bm{K}_1 \bm{K}_2 \bm{K}_3} = \bm{\Lambda}^{U,c}_{\bm{K}_1 \bm{K}_2 \bm{K}_3}$.}
\begin{align}
    \bm{\Lambda}^{U, \text{lat}}_{\bm{k}_1 \bm{k}_2 \bm{k}_3} \approx \sum_{\bm{R}_1 \bm{R}_2 \bm{R}_3} e^{i \bm{k}_1 \bm{R}_1 + i \bm{k}_2 \bm{R}_2 + i \bm{k}_3 \bm{R}_3} \bm{\Lambda}^{U, c}_{\bm{R}_1 \bm{R}_2 \bm{R}_3} \,,
    \label{eq:itp_nosym}
\end{align}
which can be solved by obtaining the coefficients $\bm{\Lambda}^{U, c}_{\bm{R}_1 \bm{R}_2 \bm{R}_3}$ from a discrete Fourier transform of $\bm{\Lambda}^{U, c}_{\bm{K}_1 \bm{K}_2 \bm{K}_3}$ onto an auxiliary real space cluster spanned by $\bm{R}$.

\begin{figure}
    \centering
    \includegraphics[width=\columnwidth]{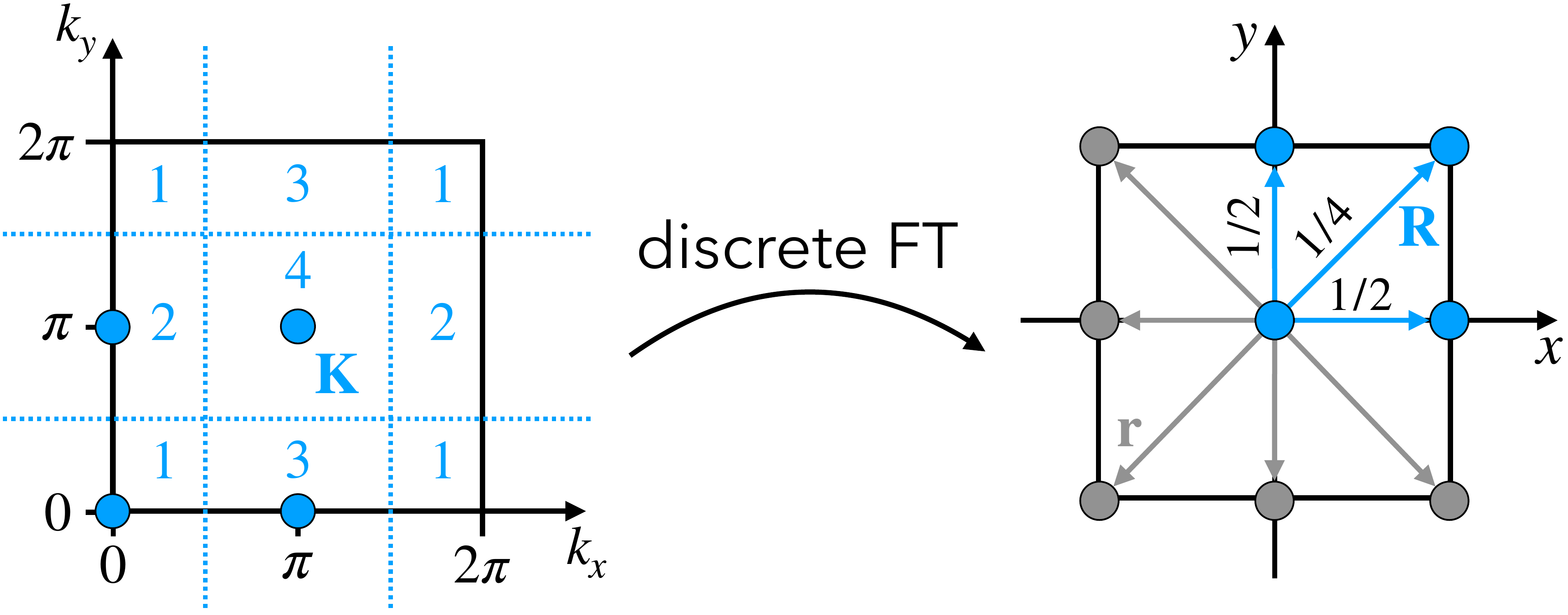}
    \caption{\textbf{Lattice parametrization of $\bm{\Lambda}^U$ in $\DCAsq$}. The irreducible vertex on the lattice is interpolated by associating a set of distance vectors $\bm{R}$ (blue arrows, right side) with the coarse-grained DCA cluster momenta $\bm{K}$ (blue dots, left side) via a discrete Fourier transform. To incorporate lattice symmetries, one subsequently includes all $\bm{r}$'s related to the $\bm{R}$'s by point group symmetry (blue and grey arrows, right side) and truncates the Fourier transform when projecting back to (continuous) momentum space.}
    \label{fig:cluster_rep}
\end{figure}

Equation \eqref{eq:itp_nosym} does, however, not guarantee the fulfillment of point group symmetries. We hence generate all $\bm{r}$ (and corresponding vertices)  that are related by point group symmetries to $\bm{R}$ (see Fig.~\ref{fig:cluster_rep})
with appropriately chosen weighting factors $c_{\bm{r}_1 \bm{r}_2 \bm{r}_3}$~\footnote{In practice, $c_{\bm{r}_1 \bm{r}_2 \bm{r}_3}$ is computed by taking the ratio between the number of bonds $(\bm{R}_1, \bm{R}_2, \bm{R}_3)$ included in the cluster and the total number of symmetry equivalent bonds for $(\bm{r}_1, \bm{r}_2, \bm{r}_3)$.} such that the symmetrized vertex $\bm{\Lambda}^{U, \text{lat}, \text{sym}}_{\bm{k}_1 \bm{k}_2 \bm{k}_3}$ coincides with the DCA vertex on the cluster momenta,
\begin{align}
    \bm{\Lambda}^{U, \text{lat,sym}}_{\bm{k}_1 \bm{k}_2 \bm{k}_3} \approx \sum_{\bm{r}_1 \bm{r}_2 \bm{r}_3} c_{\bm{r}_1 \bm{r}_2 \bm{r}_3} e^{i \bm{k}_1 \bm{r}_1 + i \bm{k}_2 \bm{r}_2 + i \bm{k}_3 \bm{r}_3} \bm{\Lambda}^U_{\bm{r}_1 \bm{r}_2 \bm{r}_3} \,.
    \label{eq:itp_sym}
\end{align}
If $\bm{\Lambda}^U$ is indeed short ranged (as we assume in eMBEX), this procedure provides us with a reasonable lattice approximation of the vertex: its Fourier series in real space must converge rapidly and the summation can be truncated at the maximum length scale of the cluster. In the following, we drop the additional superscripts (\emph{lat, sym}) for brevity and explicitly associate $\bm{\Lambda}^{U}$ on the lattice with our representation.

\section{Self-consistent SBE equations}
\label{sec:SBE_equations}
\begin{figure*}
    \centering
    \includegraphics[width=1.7\columnwidth]{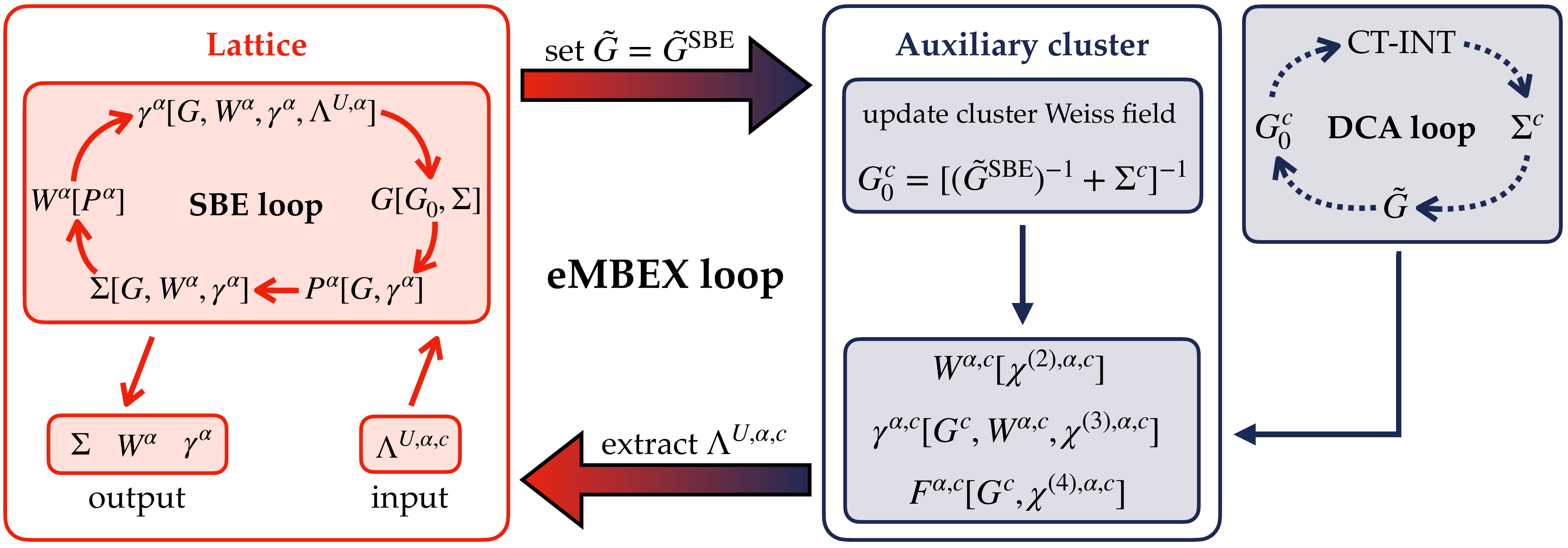}
    \caption{\textbf{Self-consistency cycle of the eMBEX method.} The algorithm is detailed step-by-step in Sec.~\ref{subsec:SBE_solver}. Note that the DCA loop for the cluster is only converged once for initialization purposes, later $\tilde{G}^{\rm SBE}$ and the cluster self-energy $\Sigma^c$ from the previous iteration is employed to update the cluster Weiss field.}
    \label{fig:embex_schematic}
\end{figure*}
With an approximation for the irreducible vertex $\bm{\Lambda}^{U}$ at hand, the quantities $\overline{\bm{\gamma}},
\bm{\gamma}, \bm{W}$ in Eq.~\eqref{eq:delta}, as well as the fermionic and bosonic self-energies defined below,
can be determined self-consistently~\cite{Krien_2019_2,Krien_2020_1}.
In this section we present the corresponding set of equations for the eMBEX method. Since they recover the SBE quantities $\bm{\gamma}$ and $\bm{W}$ from the multi-boson vertex $\bm{\Lambda}^U$, we refer to them as `SBE equations'.

We consider the SU($2$)-symmetric single-band Hubbard model, i.e.~the multi-index $\ell$ from Sec.~\ref{sec:SBE_decomposition} only incorporates spin, denoted by $\sigma$, apart from momentum and frequency arguments. For each vertex function only two spin components,
$(\sigma_1 \sigma_2 \sigma_3 \sigma_4) = (\uparrow \uparrow \uparrow \uparrow)$ and $(\sigma_1 \sigma_2 \sigma_3 \sigma_4) = (\uparrow \uparrow \downarrow \downarrow)$, need to be considered. In this case, it is convenient to introduce three physical channels, charge (ch), spin (sp), and singlet pairing (s)~\footnote{
In the single-band Hubbard model a triplet channel, $A^{\triplet}=\bm{A}^{\uparrow \uparrow \uparrow \uparrow}$,
is not required for the SBE decomposition, since the bare Hubbard interaction acts only between spin of opposite flavor.}.
For an arbitrary four-point function $\bm{A}$, the physical channels are defined as linear combinations of the remaining spin components,
\begin{subequations}
    \begin{align}
        A^{\charge} &= \bm{A}^{\uparrow \uparrow \uparrow \uparrow} + \bm{A}^{\uparrow \uparrow \downarrow \downarrow}, \label{eq:ch}\\ 
        A^{\spin} &= \bm{A}^{\uparrow \uparrow \uparrow \uparrow} - \bm{A}^{\uparrow \uparrow \downarrow \downarrow}, \label{eq:sp}\\ 
        A^{\singlet} &= 2 \bm{A}^{\uparrow \uparrow \downarrow \downarrow} - \bm{A}^{\uparrow \uparrow \uparrow \uparrow} \label{eq:si}\,.
    \end{align}
\end{subequations}
The physical flavors `$\charge$' and `$\spin$' are associated to the particle-hole channel, and the `$\singlet$' flavor to the particle-particle channel~\footnote{The `$\charge$'/`$\spin$' (`$\singlet$'/`$\triplet$')
notation diagonalizes the Bethe-Salpeter equation in the particle-hole (particle-particle) channel~\cite{Rohringer_2012,Bickers_2004}.}.
Therefore, after moving to the `$\charge$', `$\spin$', `$\singlet$' notation, we drop the labels `$\ph$', `$\cph$', and `$\pp$' where applicable.
The SBE decomposition~\eqref{eq:sbe_decomposition} is explicitly shown in this notation in Appendix~\ref{app:sbe_chsps}.

We further define fermionic and bosonic momentum-frequency four-vectors,
\begin{align}
    k=(\bm{k},\nu),\;\;\;q=(\bm{q},\omega),
\end{align}
where $\nu$ and $\omega$ are corresponding fermionic/bosonic Matsubara frequencies. Summations $\sum_k$, $\sum_q$ carry an implicit factor $T/N_{\bm{k}}$, where $T$ is the temperature and $N_{\bm{k}}$ the total number of lattice sites.

The Hedin vertex of flavor `$\charge$', `$\spin$', `$\singlet$' is a three-leg vertex that is $U$-irreducible in the corresponding channel,
\begin{subequations}
    \begin{align}
        \gamma^{\charge/\spin}_{kq}=1+\sum_{k'}(F^{\charge/\spin}_{kk'q}-\Delta^{\charge/\spin}_{kk'q})G_{k'}G_{k'+q}\label{eq:hedin_ph_f},\\
        \gamma^{\singlet}_{kq}=1-\frac{1}{2}\sum_{k'}(F^{\singlet}_{kk'q}-\Delta^{\singlet}_{kk'q})G_{k'}G_{q-k'}\label{eq:hedin_pp_f} \,,
    \end{align}
\end{subequations}
where $F^{\alpha} - \Delta^{\alpha}$ with $\alpha \in \{\charge/\spin, \singlet \}$ depends on the multi-boson vertex $\Lambda^{U, \alpha}$ through the SBE decomposition in Eq.~\eqref{eq:sbe_decomposition} as is explicitly shown in Appendix~\ref{app:sbe_chsps}.

The Hedin vertex determines the fermionic self-energy $\Sigma$ and the polarization $\pol$, i.e.~the bosonic self-energy. In principle, the expression for the self-energy is subject to the Fierz ambiguity
of the Hubbard interaction~\cite{Ayral_2016,Ayral_2017}, cf.~Appendix~\ref{app:fierz}. However, we show in Appendix~\ref{app:fierz_sym} that the Fierz ambiguity is resolved because the input of the calculation, $\Lambda^U$, satisfies the crossing symmetry~\cite{Rohringer_2012}. The self-energy in Eq.~\eqref{eq:hedin_fierz} of the Appendix thus does not depend on the Fierz parameter and we have without loss of generality:
\begin{align}
    \Sigma_k=& \Sigma_{\text{H}}-\frac{1}{2}\sum_qG_{k+q}\left[W^\charge_q\gamma^\charge_{kq}+W^\spin_q\gamma^\spin_{kq}\right],\label{eq:hedin}
\end{align}
where $\Sigma_{\text{H}}= \tfrac{U}{2} \langle n \rangle$ denotes the Hartree shift with
$\langle n\rangle=2\sum_k G_k$ being the average occupation per site
(a convergence factor is implicitly assumed in the sum over Matsubara frequencies). By virtue of the two-particle self-consistency of our approach,
the $W$'s in Eq.~\eqref{eq:hedin} contain themselves the crossing-symmetric vertex corrections,
which ensures that the exact leading asymptotic coefficient of the self-energy is recovered~\cite{Kugler_2022}.

The polarization in the various channels is given as,
\begin{subequations}
        \begin{align}
        \pol_q^{\charge/\spin}=&\sum_k G_k G_{k+q}\gamma^{\charge/\spin}_{kq},\label{eq:polarization_chsp}\\
        \pol_q^{\singlet}=&-\sum_k G_k G_{q-k}\gamma^{\singlet}_{kq}\label{eq:polarization_sing}.
    \end{align}
\end{subequations}
Lastly, the fermionic and bosonic propagators $G$ and $W$ are obtained via the Dyson equations,
\begin{align}
    G_k=&\frac{G_{0,k}}{1-G_{0,k}\Sigma_k},\label{eq:dyson}
\end{align}
where $G_{0,k}=\left[i\nu+\mu-\varepsilon_{\bm{k}}\right]^{-1}$ is the bare Green's function,
$\mu$ the chemical potential, $\varepsilon_{\bm{k}}$ the noninteracting dispersion, and
\begin{subequations}
    \begin{align}
        W_q^{\charge/\spin}=&\frac{\pm U}{1\mp U\pol_q^{\charge/\spin}},\label{eq:dyson_chsp}\\
        W_q^{\singlet}=&\frac{2U}{1+U\pol_q^{\singlet}}.\label{eq:dyson_sing}
    \end{align}
\end{subequations}
As discussed in Refs.~\cite{Krien_2019_2, Gievers_2022, Kiese_2024}, Eqs.~\eqref{eq:hedin_ph_f} to~\eqref{eq:dyson_sing}
are a closed set of equations for given input $\Lambda^U$.
The corresponding calculation cycle is shown in Fig.~\ref{fig:embex_schematic}.

\section{Implementation}\label{sec:implementation}

\subsection{Dynamical cluster approximation}
\label{subsec:DCA}

In more detail, we consider in DCA \cite{Hettler_1998} an auxiliary model of interacting electrons $f$, localized on a periodic cluster of $N_c$ sites
\begin{align}
    S_c[\bar{f}, f] = -&\int_0^{\beta}\!\!d\tau d\tau' \sum_{ij, \sigma} \bar{f}_{i \sigma}(\tau) [G^{\sigma, c}_{0}]^{-1}_{ij}(\tau - \tau') f_{j \sigma}(\tau') \notag \\ 
    + &\int_0^{\beta} d\tau \sum_i U \bar{f}_{i \uparrow}(\tau) f_{i \uparrow}(\tau) \bar{f}_{i \downarrow}(\tau) f_{i \downarrow}(\tau) \,,
    \label{eq:cluster_model}
\end{align}
and use the DCA approximation
\begin{align}
    \Sigma^{\sigma}_k \approx \sum_{\bm{K}} \Theta_{\bm{K}}(\bm{k}) \Sigma^{\sigma, c}_{K} \,,
\end{align}
for the lattice self-energy. The function $\Theta$ is only finite, i.e.~$\Theta_{\bm{K}}(\bm{k}) = 1$, if $\bm{k}$ is located in the patch associated with the cluster momentum $\bm{K}$. We compute the cluster self-energy $\Sigma^{\sigma, c}$ using the CT-INT quantum Monte Carlo solver implemented in the TRIQS library \cite{Parcollet_2015}. Details on the calculation of multi-point correlation functions within CT-INT can be found in Ref.~\cite{Ayral_thesis}. We omit spin labels $\sigma$ for the remainder of this section, since we assumed spin-rotation invariance when discussing the SBE equations. The Weiss field is given by
\begin{align}
    G^c_{0, K} = \left[ \tilde{G}^{-1}_K + \Sigma^{c}_K \right]^{-1} \,,
    \label{eq:DCA_bath_update}
\end{align}
where $\tilde{G}$ is the coarse grained lattice Green's function 
\begin{align}
    \tilde{G}_K = \frac{N_c}{N_{\bm{k}}} \sum_{\bm{k}} \Theta_{\bm{K}}(\bm{k}) G_k \,,
\end{align}
and $N_{\bm{k}}$ denotes the number of lattice sites. In practice, one starts with an initial guess for the cluster self-energy, e.g.~$\Sigma^c = 0$, calculates $G^c_0$ from Eq.~\eqref{eq:DCA_bath_update} and solves the cluster model to obtain an updated $\Sigma^c$. This process is iterated until the self-consistency condition $G^c = \tilde{G}$ is fulfilled to sufficient numerical accuracy~\footnote{Note that the SBE equations would recover the DCA solution if the diagrammatic calculation is started from the cluster vertex $\Lambda^{U, c}$ \emph{and} the cluster Weiss field $G^c_0$.}.

\subsection{SBE solver}
\label{subsec:SBE_solver}
\begin{figure}
    \centering
    \includegraphics[width = \columnwidth]{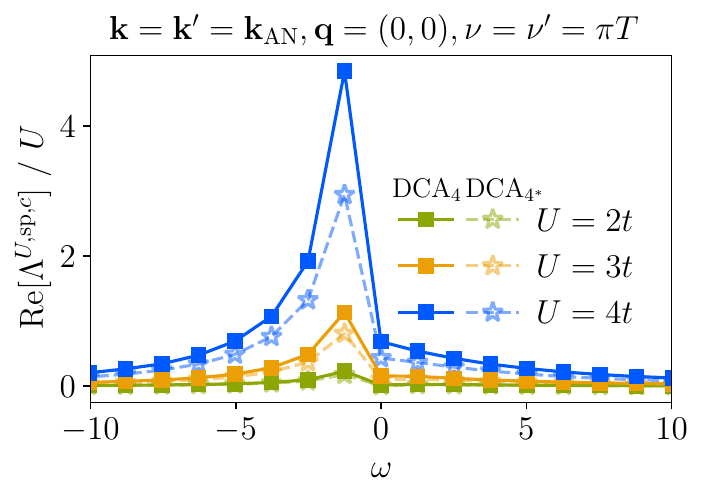}
    \caption{\textbf{Low-frequency structure of $\Lambda^{U, c}$ from DCA}. We show the real part of $\Lambda^{U, \spin, c}_{\bm{k} \bm{k}' \bm{q} \nu \nu' \omega}$ in $\DCAsq$ (squares) and $\DCAst$ (stars) for various values of $U$ and as a function of transfer frequency $\omega$. The values of all other arguments are fixed  such that we slice through the vertex maximum.}
    \label{fig:Lambda_U_Omega_slice}
\end{figure}
The eMBEX cycle is schematically summarized in Fig.~\ref{fig:embex_schematic} and consists of the following steps:
\begin{enumerate}[label = (\arabic*)]
    \item Converge the DCA loop to find the lattice Green's function $G = G^{\mathrm{DCA}}$ as detailed in Sec.~\ref{subsec:DCA}.
    \item At convergence, measure the two, three and four-point correlation functions ($\chi^{(2), \alpha, c}$, $\chi^{(3), \alpha, c}$, $\chi^{(4), \alpha, c}$) \cite{Krien_2020_1} and compute $W^{\alpha, c}$, $\gamma^{\alpha, c}$ and $F^{\alpha, c}$ for the cluster ($\alpha \in \{\charge, \spin, \singlet \}$).
    \item Calculate the interaction irreducible cluster vertex $\Lambda^{U, \alpha, c}$ using the cluster analog of Eq.~\eqref{eq:sbe_decomposition}. Periodize according to Eq.~\eqref{eq:itp_sym} to obtain the irreducible vertex on the lattice.
    \item Initialize the self-energy, screened interactions and Hedin vertices for the self-consistent SBE calculations on the lattice. One possible choice is $\Sigma = \Sigma_{\text{H}}$, $W^{\charge/ \spin} = \pm U$, $W^{\singlet} = 2U$ and $\gamma^{\alpha} = 1$.
    \item Calculate $G$ from the Dyson equation~\eqref{eq:dyson}. 
    \item Update the polarization $\pol^{\alpha}$ using Eqs.~\eqref{eq:polarization_chsp} \& \eqref{eq:polarization_sing}.
    \item Use the equation of motion \eqref{eq:hedin} to compute $\Sigma$.
    \item Determine the Hedin vertices and screened interactions through Eqs.~\eqref{eq:hedin_ph_f}, \eqref{eq:hedin_pp_f} and Eqs.~\eqref{eq:dyson_chsp}, \eqref{eq:dyson_sing}.
    \item Repeat steps (5) - (8) until the solution is converged to sufficient numerical accuracy. We terminate the SBE loop once the relative deviation $|| \tilde{x} - x || / ||x||$ between the solution vectors of the current and previous iteration ($\tilde{x}$ and $x$, respectively) becomes smaller than $\epsilon = 10^{-3}$. Here, $||.||$ denotes the $\infty$-norm and $x = (\Sigma, W^{\alpha}, \gamma^{\alpha})$.
    \item Update the bath for the cluster model using Eq.~\eqref{eq:DCA_bath_update}, replacing $\tilde{G}$ with the coarse-grained lattice Green's function from SBE.
    \item Repeat (2) - (10) until the outer self-consistency condition $G^c = \tilde{G}^{\mathrm{SBE}}$ is met.
\end{enumerate}
When iterating steps (5) - (8), we empirically find that the Dyson equations for the screened interactions $W^{\alpha}$ [cf.~Eqs.~\eqref{eq:dyson_chsp} \& \eqref{eq:dyson_sing}] are numerically unstable due to possible poles on the right-hand side of the equation. Therefore, we use the iterative equations
\begin{align}
    W_{q, n + 1}^{\charge/\spin} &= \pm U \pm U W_{q, n}^{\charge/\spin} \pol_{q, n}^{\charge/\spin} \\
    W_{q, n + 1}^{\singlet} &= 2U - U W_{q, n}^{\singlet} \pol_{q, n}^{\singlet} \,,
\end{align}
where $n$ is the iteration index. To determine the SBE fixed point, we employ the periodic Pulay mixing scheme of Ref.~\cite{Banerjee_2016}. 

\section{Results}\label{sec:results}
\begin{figure}
    \centering
    \includegraphics[width = \linewidth]{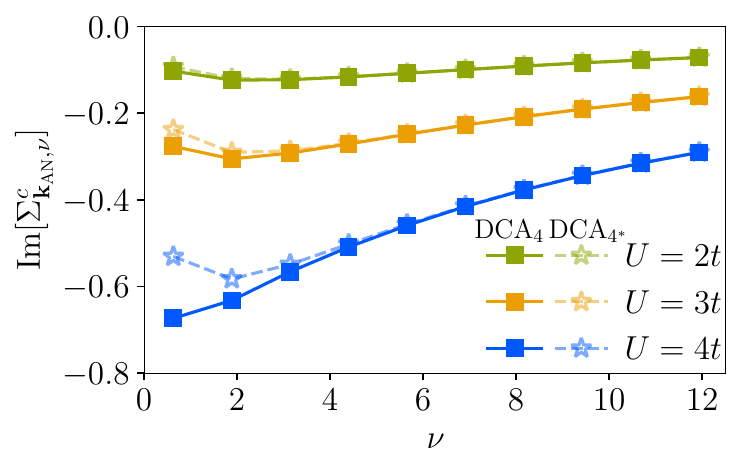}
    \caption{\textbf{Cluster self energy from DCA at the antinode}. We show results for $\DCAsq$ (squares) and $\DCAst$ (stars) at $\bm{k}_{\text{AN}} = (\pi, 0)$ for the first 10 Matsubara frequencies.}
    \label{fig:sigma_DCA}
\end{figure}
We consider the Hubbard model on the square lattice,
\begin{align}
    H = -t \sum_{\langle ij \rangle \sigma} c^{\dagger}_{i \sigma} c^{\phantom{\dagger}}_{j \sigma} + U \sum_i n_{i \uparrow} n_{i \downarrow} - \mu \sum_i (n_{i \uparrow} + n_{i \downarrow}) \,,
\end{align}
where $t$ is the nearest-neighbor hopping amplitude, $c$ ($c^{\dagger}$) denotes the annihilation (creation) operator, and $n_\sigma$ the density operator of electrons with spin $\sigma=\{\uparrow,\downarrow\}$.

We fix the average filling $\langle n\rangle=\langle n_\uparrow\rangle+\langle n_\downarrow\rangle$ to one electron per site and perform eMBEX calculations at a fixed inverse temperature $\beta t = 5$ for various values of $U \leq 4t$. Our SBE calculations are carried out on a 26$\times$26 momentum space grid with 16 positive fermionic and 24 positive (including $\omega = 0$) bosonic Matsubara frequencies for the Hedin vertices. The frequency boxes for all other quantities (the Green's function, polarizations etc.) are chosen much larger ($> 128$ positive frequencies) due to their reduced memory footprint and lower computational cost; outside of its smaller frequency box the Hedin vertex is approximated as $\gamma\approx1$. We ensured that our results do not change significantly when increasing the momentum resolution or the box sizes.

\begin{figure}
    \centering
    \includegraphics[width = \columnwidth]{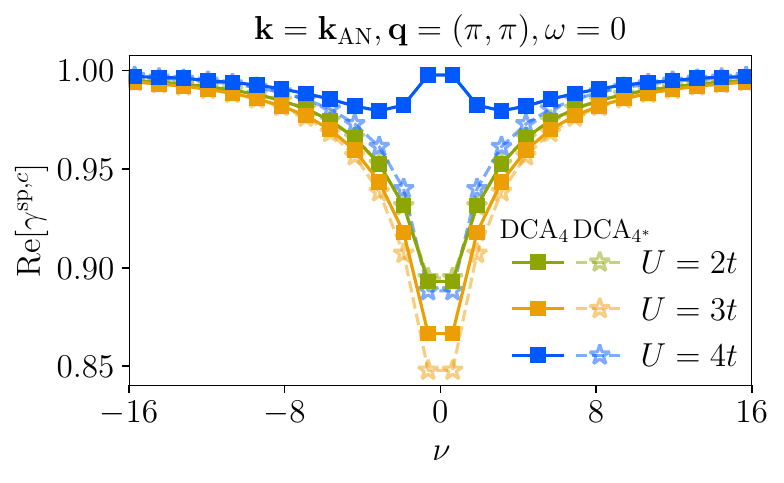}
    \caption{\textbf{Low-frequency structure of $\gamma^c$ from DCA}. We show the real part of $\gamma^{\spin, c}_{\bm{k} \bm{q} \nu \omega}$ in $\DCAsq$ (squares) and $\DCAst$ (stars) for various values of $U$ and as a function of fermionic frequency $\nu$.}
    \label{fig:gamma_DCA}
\end{figure}

\begin{figure*}
    \centering
    \includegraphics[width = \linewidth]{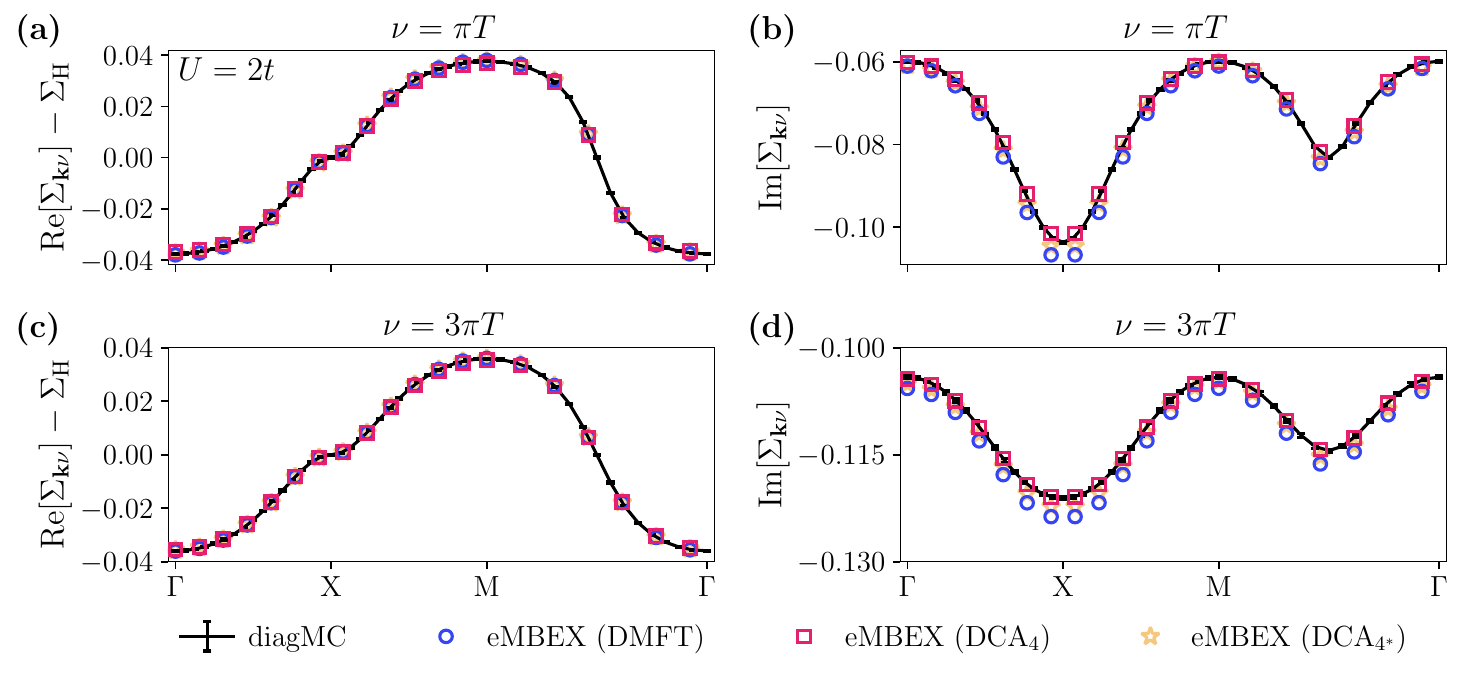}
    \caption{\textbf{Benchmark results for the half-filled Hubbard model at $U = 2t$ and $\beta t = 5$}. We show the real (left column) and imaginary part (right column) of the self-energy at the first (top row) and second Matsubara frequency (bottom row) from eMBEX with DMFT, $\DCAsq$ and $\DCAst$ input together with results from diagrammatic Monte Carlo simulations. Note that, although the self-consistent SBE equations have been solved on a 26$\times$26 lattice, we only show every second data point for clarity.}
    \label{fig:sigma_benchmark_U2}
\end{figure*}

As explained in Sec.~\ref{sec:parametrization}, we investigate single-site DMFT and two different patching schemes of the 4-site DCA as the starting point for eMBEX. In the following, we first discuss the square patching $\DCAsq$ (denoted with squares in Figs.~\ref{fig:Lambda_U_Omega_slice}-\ref{fig:sigma_benchmark_U4}) and DMFT (circles). In Sec.~\ref{subsec:reference_system} we then compare to the star patching $\DCAst$, shown as faint stars in the figures.

We focus on \emph{single-shot} calculations using the Weiss field of DMFT/DCA, that is,
the outer self-consistency loop is omitted (see large right arrow in Fig.~\ref{fig:embex_schematic}).
The role of the Weiss field is discussed in Secs.~\ref{subsec:outer_sc} and~\ref{subsec:reference_system}.

\subsection{Input from DCA and lattice parametrization}

We begin with a discussion of the cluster correlation functions corresponding to the $\DCAsq$. The central object of our method is the irreducible cluster vertex $\Lambda^{U,c}$.
As it depends on three momenta and frequencies, we focus on a one-dimensional
cut along the bosonic $\omega$-direction with all other arguments fixed.
As can be seen from Fig.~\ref{fig:Lambda_U_Omega_slice}, $\Lambda^{U,c}$ decays away from $\omega=0$,
where it shows a pronounced peak that rapidly increases as a function of $U/t$.

Fig.~\ref{fig:sigma_DCA} shows the evolution of the imaginary part of the
antinodal cluster self-energy from metallic ($U/t=2$) to insulating ($U/t=4$),
as roughly indicated by the negative/positive slope of $\text{Im}\Sigma^c_{\bm{k}_{\text{AN}},\nu}$ at the smallest Matsubara frequencies.

A further indication of this change of regime is visible in the low-frequency behavior of
the spin-fermion coupling $\gamma^{\spin, c}$ of the cluster shown in Fig.~\ref{fig:gamma_DCA}.
We focus on the antinodal fermionic momentum $\bm{k}_{\text{AN}}$ and on the bosonic momentum $\bm{q}=(\pi,\pi)$,
that is, the coupling of antinodal fermions to static ($\omega=0$) antiferromagnetic spin fluctuations.
While for $U/t=2$ and $3$ the real part of $\gamma^{\spin, c}$ is suppressed ($<1$),
see green and yellow squares in Fig.~\ref{fig:gamma_DCA}, it is enhanced in the insulating regime $U/t=4$ (blue).
The enhancement is selective with respect to the fermionic momentum $\bm{k}$ and we observe it only at the antinode,
whereas for other cluster momenta the vertex is still suppressed for small $\nu$ (not shown).

The suppression of $\gamma^{\spin, c}$ for weak coupling is the result of Kanamori screening,
corresponding to a particle-particle vertex correction~\cite{Krien_2020_1,Kitatani2019}.
The enhancement at $U/t=4$, on the other hand, is reminiscent of the behavior of the momentum-independent
spin-fermion coupling of DMFT in the bad-metallic and Mott-insulating regimes~\cite{Katanin2009,Harkov_2021}.
There, it appears simultaneously with the formation of the local moment~\cite{Adler_2024},
which marks the onset of the non-perturbative regime~\cite{Chalupa_2021}.

\subsection{Self-energy benchmarks}
\label{subsec:benchmarks}
\begin{figure*}
    \centering
    \includegraphics[width = \linewidth]{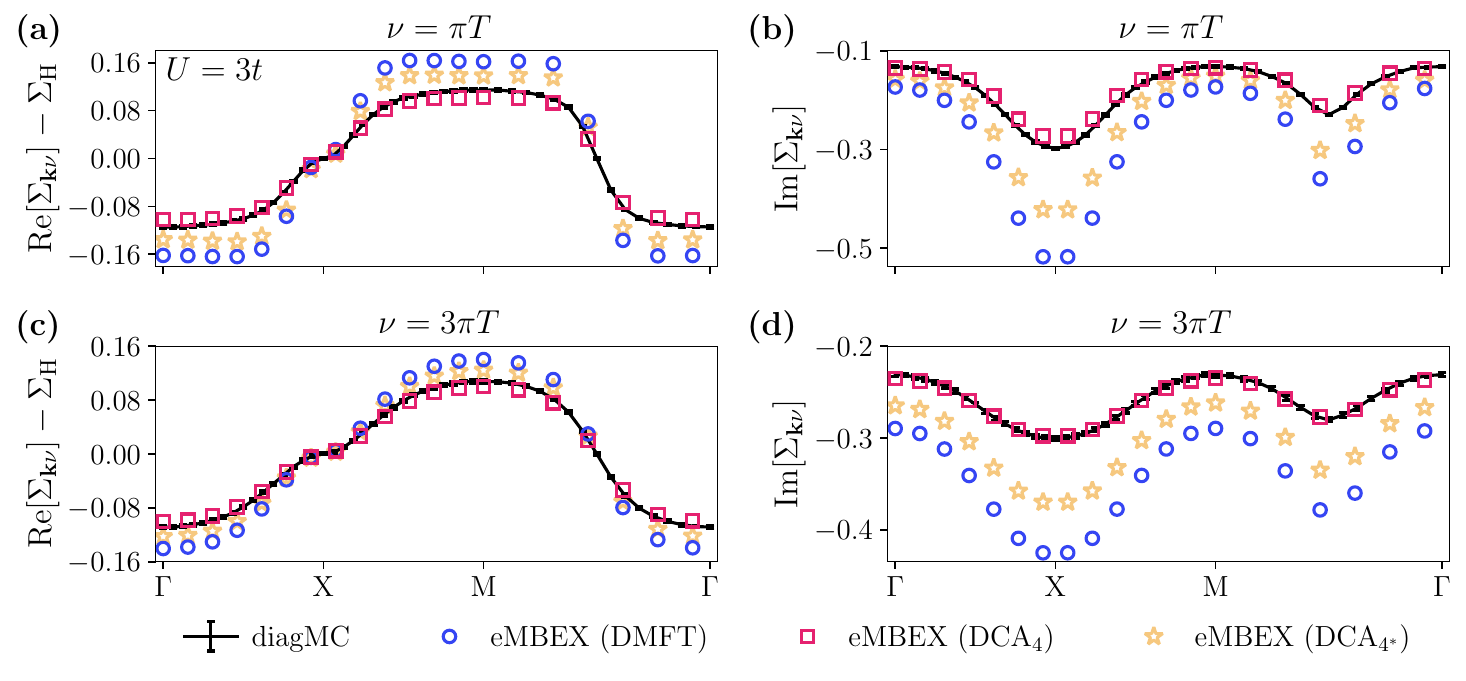}
    \caption{\textbf{Benchmark results for the half-filled Hubbard model at $U = 3t$ and $\beta t = 5$}. Same as Fig.~\ref{fig:sigma_benchmark_U2}, except for $U$.}
    \label{fig:sigma_benchmark_U3}
\end{figure*}
In order to gauge the quantitative accuracy of the eMBEX approach with DMFT and $\DCAsq$ input, we compare the converged eMBEX lattice self-energy at the first two Matsubara frequencies to results obtained with
numerically exact diagrammatic Monte Carlo (diagMC) simulations for $U/t=2, 3$, and $4$~\cite{Klett_2020}.

For $U = 2t$, we find that the real part of the self-energy [see Fig.~\ref{fig:sigma_benchmark_U2}(a) \& (c)]
is in very good agreement with diagMC.
The imaginary part [Fig.~\ref{fig:sigma_benchmark_U2}(b) \& (d)]
shows slight deviations for the DMFT input (blue circles) near the anti-node $\bm{k}_{\text{AN}} = (\pi, 0)$
and node $\bm{k}_{\text{N}} = (\tfrac{\pi}{2}, \tfrac{\pi}{2})$, as well as a small offset. Fig.~\ref{fig:sigma_benchmark_U3} shows that for larger $U = 3t$, the agreement between single-site eMBEX and diagMC detoriates further, while the $\DCAsq$ shows an excellent agreement for both the real and imaginary part of $\Sigma$.

For $U/t = 4$ we were not able to converge eMBEX based on DMFT
and the convergence of the self-consistent cycle using $\DCAsq$ input is substantially more difficult. In the latter case, we achieved convergence only on a smaller $6\times6$ lattice due to the increased computational overhead of invoking a linear preconditioner~\cite{Baker_2005, Cai_2002}. For this reason our benchmark for this value of $U$ is discussed separately in Appendix~\ref{app:U4_benchmark}.

\subsection{Screened interaction}

\begin{figure*}
    \centering
    \includegraphics[width = \linewidth]{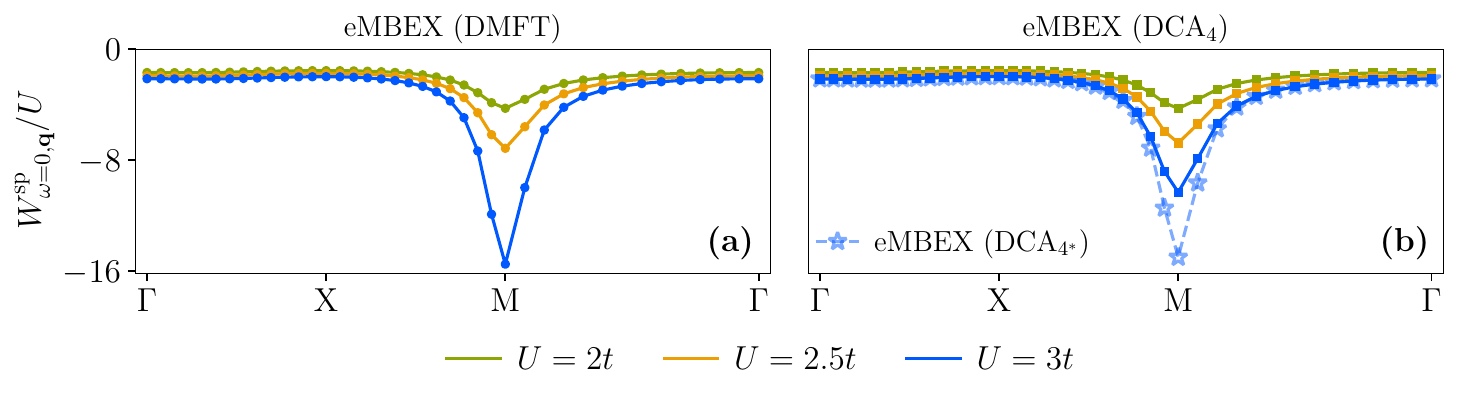}
    \caption{\textbf{eMBEX result for $W^{\spin}$ for both DMFT input (left panel) and DCA input (right panel)}. Antiferromagnetic fluctuations are damped when increasing the cluster size as indicated by a reduced peak height around $M = (\pi, \pi)$. For reference, we also show the $\DCAst$ result at $U = 3t$ in (b).}
    \label{fig:Wsp_k}
\end{figure*}

At half filling, static spin fluctuations constitute the dominant bosonic mode. Therefore,
we study the evolution of $W^{\spin}_{\bm{q},\omega=0}$ with $U$ in Fig.~\ref{fig:Wsp_k} for the DMFT and $\DCAsq$ inputs.
As expected, $W^{\spin}$ is almost featureless in momentum space
except for a pronounced peak at $\bm{q} = (\pi, \pi)$, see also, for example, Ref.~\cite{Krien_2022_2}.

At weak coupling, $U \lesssim 2.5t$, the height of the peak grows
only moderately with $U$ both for DMFT (left panel of Fig.~\ref{fig:Wsp_k}) and $\DCAsq$ (right panel) input.
For $U = 3t$, however, single-site eMBEX strongly overestimates the peak height (compare dark blue lines in left and right panel).
As a result, the feedback of antiferromagnetic correlations on the self-energy is too strong,
which explains the overestimation of single-particle correlations observed in Fig.~\ref{fig:sigma_benchmark_U3}.
Indeed, DMFT overestimates magnetic correlations~\cite{Ayral_2015} and we argue in Appendix~\ref{app:effective_exchange}
that a cluster size of $N_c>1$ is mandatory in practical applications.

\subsection{Three-leg vertex}

\begin{figure*}
    \centering
    \includegraphics[width = \linewidth]{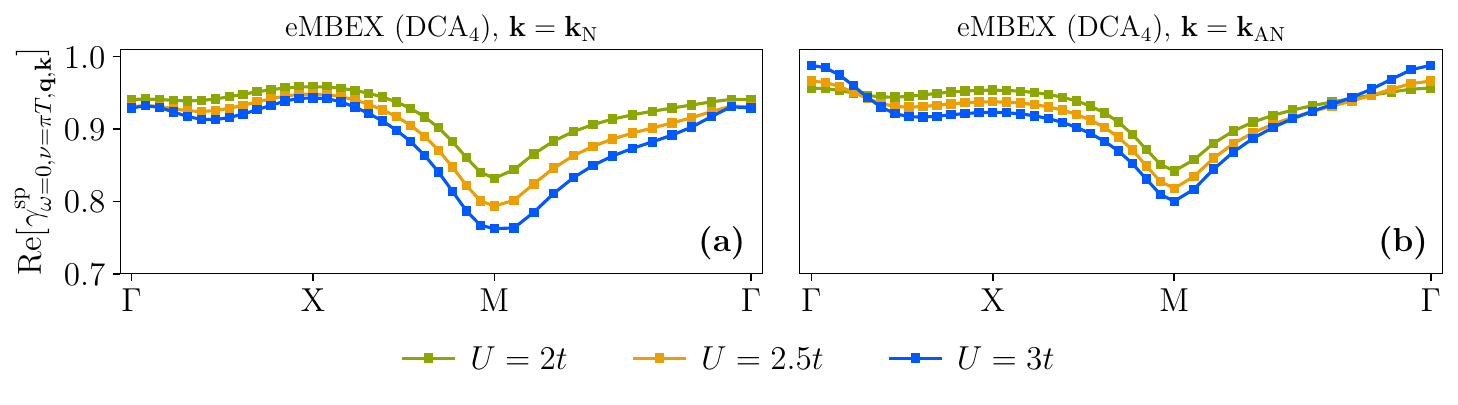}
    \caption{\textbf{$\gamma^{\spin}$ from eMBEX with $\DCAsq$ input for various $U$}. We show the low-energy spin-fermion coupling for nodal (left panel) and antinodal (right panel) fermions as a function of the bosonic momentum $\bm{q}$.}
    \label{fig:gamma_sp_k}
\end{figure*}

In Fig.~\ref{fig:gamma_DCA} we already examined the spin-fermion coupling of the $\DCAsq$ cluster.
In the following, we examine the spin-fermion coupling $\gamma^{\spin}_{\bm{k} \bm{q} \nu \omega}$
defined on the large $26\times26$ lattice, shown in Fig.~\ref{fig:gamma_sp_k}, 
which is computed during the eMBEX cycle via equations~\eqref{eq:hedin_ph_f} and~\eqref{eq:hedin_pp_f}.
For $U/t \leq 3$, where a converged eMBEX solution can easily be obtained, $\text{Re}\gamma^\spin$ is suppressed
below its noninteracting value $1$.

However, for increasing interaction the coupling to ferromagnetic spin fluctuations $\bm{q}=(0,0)$
is enhanced for antinodal fermions, e.g., $\bm{k}_{\text{AN}}=(\pi,0)$, see the right panel in Fig.~\ref{fig:gamma_sp_k}.
Indeed, in the atomic limit the vertex is enhanced at low temperature~\cite{Harkov_2021},
and considering the Hubbard model from this strong-coupling limit it is plausible
that this vertex changes from being suppressed to being enhanced with increasing interaction.
The right panel of Fig.~\ref{fig:gamma_sp_k} suggests that this evolution could be momentum selective,
beginning with antinodal fermions coupled to ferromagnetic correlations.

\subsection{Outer self-consistency}
\label{subsec:outer_sc}

We tentatively discuss the possibility of implementing an \emph{outer} (one-particle) self-consistency in eMBEX by updating the cluster Weiss field from a coarse graining of the SBE Green's function (see Sec.~\ref{subsec:DCA}). We here restrict ourselves to a single iteration, because of the high computational cost associated with the repeated calculation of the irreducible cluster vertex $\Lambda^U$. Figure $\ref{fig:bath_update}$ compares the bath for the embedded cluster before and after the SBE update (the insets in (c) and (d) show the absolute deviation). We find that the cluster and eMBEX Weiss fields for $\DCAsq$ are almost identical for $U/t = 2, 3$. This suggests that the original Weiss field of $\DCAsq$ is a suitable choice in the considered regimes.

\subsection{Relevance of the cluster reference system}
\label{subsec:reference_system}

To further elucidate the impact of the cluster reference system, we now turn to eMBEX calculations based on the alternative four-site
star patching $\DCAst$ shown on the right of Fig.~\ref{fig:patching_geometries} and compare to the square patching $\DCAsq$.

We observe a substantial deviation from $\DCAsq$ already on the level of the cluster correlation functions shown in
Figs.~\ref{fig:Lambda_U_Omega_slice},~\ref{fig:sigma_DCA}, and~\ref{fig:gamma_DCA}.
For each value of $U$, the irreducible vertex $\Lambda^{U,c}$ is smaller in $\DCAst$ (see stars in Fig.~\ref{fig:Lambda_U_Omega_slice}) compared to $\DCAsq$. Further, in $\DCAsq$ short-ranged spin fluctuations have a much bigger impact on the self-energy $\Sigma^c$, even forming a pseudogap for $U/t=4$, whereas the self-energy of $\DCAst$ remains metallic, see blue symbols in Fig.~\ref{fig:sigma_DCA}. In fact, we observed small deviations from perfect particle-hole symmetry in $\DCAst$ (not shown). Similarly, in $\DCAst$ the spin-fermion coupling $\gamma^{\spin, c}$ does not show the enhancement at the antinode
for $U/t=4$ that we observed in $\DCAsq$, see Fig.~\ref{fig:gamma_DCA}. All these aspects suggest that $\DCAst$ corresponds to a much less correlated
and, hence, qualitatively different starting point for eMBEX compared to $\DCAsq$.

\begin{figure}
    \centering
    \includegraphics[width = \linewidth]{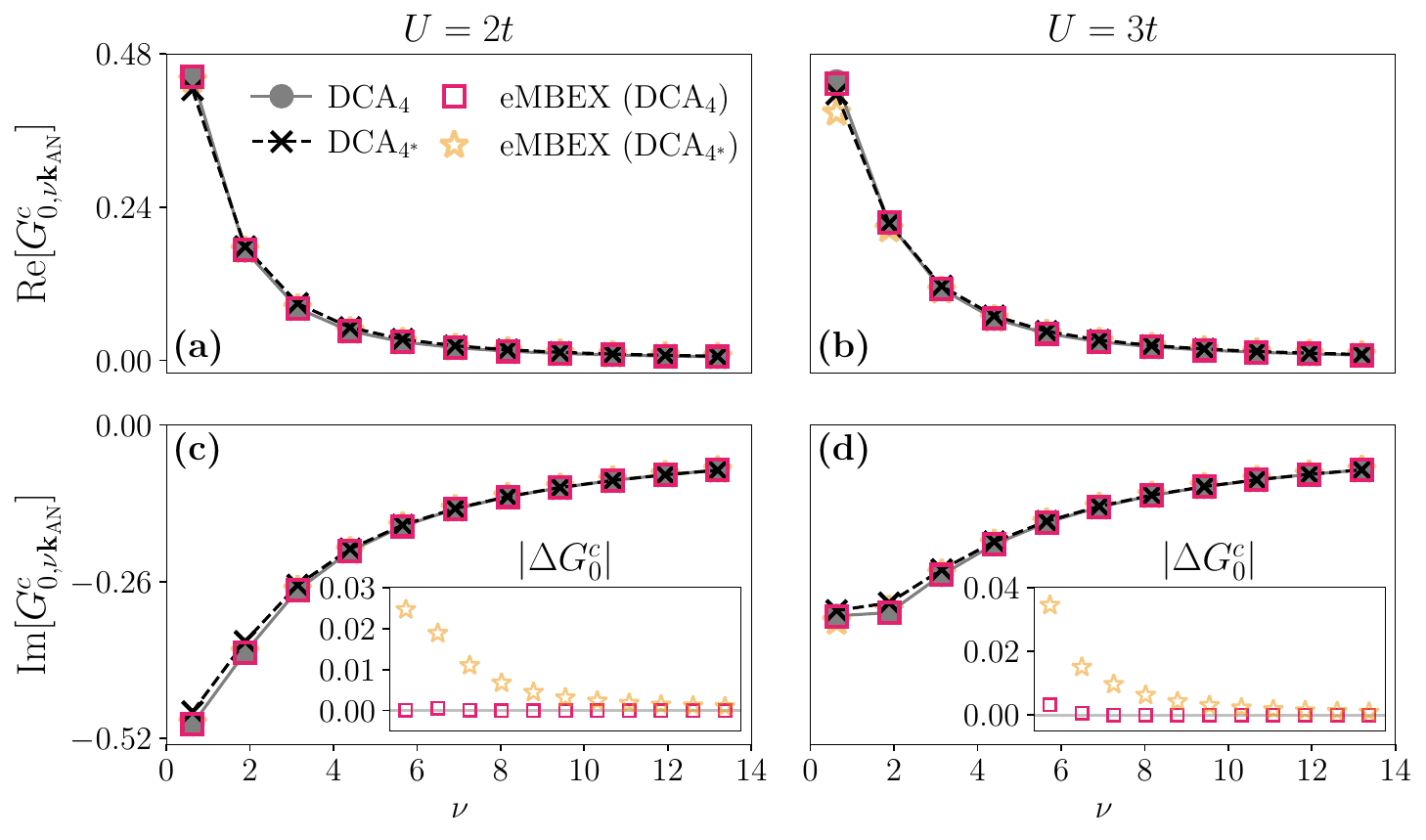}
    \caption{\textbf{Cluster Weiss field from DCA and eMBEX}. The top (bottom) row shows the real (imaginary) part at the anti-node $\bm{k}_{\text{AN}}$ for the first 10 Matsubara frequencies. The insets in (c) \& (d) show the absolute deviation between the DCA and eMBEX Weiss fields for the same choice of momentum and frequency.}
    \label{fig:bath_update}
\end{figure}

Next, we compare the eMBEX results obtained from the $\DCAsq$ and $\DCAst$ inputs.
The right panels of Fig.~\ref{fig:sigma_benchmark_U2} show already a slightly worse performance of $\DCAst$ for $\text{Im}\Sigma$ at $U/t=2$.
For $U/t=3$ (see Fig.~\ref{fig:sigma_benchmark_U3}) the self-energy differs drastically from the benchmark and is only slightly better than the single-site input from DMFT. In fact, long-ranged spin fluctuations are strongly overestimated in the eMBEX calculation based on $\DCAst$ (cf. stars in right panel of Fig.~\ref{fig:Wsp_k}). It is plausible that $\DCAst$ does not sufficiently account for short-ranged correlations. The result is, perhaps counter-intuitively, that long-ranged correlations are overestimated and their overall feedback on the self-energy is too large, similar to single-site eMBEX. This interpretation is further supported by the updated Weiss fields shown in Fig.~\ref{fig:bath_update}, whose deviation from the original Weiss fields of $\DCAst$ is much larger compared to the updated Weiss fields of $\DCAsq$.

Finally, we note that the less correlated starting point of $\DCAst$ leads to a better
convergence of the eMBEX cycle compared to $\DCAsq$, see also Appendix~\ref{app:U4_benchmark}.
It therefore appears that the important strong short-ranged correlations accounted for in $\DCAsq$
are intrinsically difficult to include into our diagrammatic scheme.
The latter corresponds to a subset of the parquet diagrams~\cite{Krien_2021} and therefore it is plausible that a similar problem
could arise in cluster extensions of the parquet D$\Gamma$A~\cite{Toschi_2007}.

\section{Conclusions}
\label{sec:conclusion}

We presented and implemented a diagrammatic multi-scale method (eMBEX), which is based
on the interaction-irreducible vertex of a quantum cluster embedded in a noninteracting medium.
Through its diagrammatic construction, eMBEX satisfies the crossing symmetry on the two-particle level and we show in Appendix~\ref{app:fierz_sym} that, as a result, the Fierz ambiguity of the Hubbard interaction is resolved exactly.

Based on four-site DCA calculations for the half-filled Hubbard model on the square lattice, we obtain results for the electronic self-energy that are in excellent agreement with the exact benchmark data. The calculations are performed in the weak-to-intermediate coupling regime,
where the antiferromagnetic correlation length exceeds the size of the cluster reference system, underlining the multi-scale feature of our approach.

We investigated two patching schemes for the four-site DCA, a simple square patching ($\DCAsq$) and a star patching ($\DCAst$), cf. Fig.~\ref{fig:patching_geometries},
where the former performs by far better in the quantitative sense. Nonetheless, we found value in the $\DCAst$ patching scheme as a qualitatively different physical starting point. In their comparison the two patching schemes show that our method relies on a proper account of strong short-ranged correlations on the level of the cluster reference system.
Interestingly, the underestimation of short-ranged correlations in $\DCAst$ leads to an overestimation of long-ranged correlations in the lattice approximation, pointing to a delicate balance between the two.

At the same time, we observed that the more strongly correlated starting point of $\DCAsq$ makes it more difficult~\cite{Tandetzky_2015} and costly to converge the self-consistent eMBEX cycle. Nevertheless, the very good agreement of four-site eMBEX with numerically exact benchmarks serves as a strong incentive to investigate more sophisticated approaches to find the fixed point. One could, for example, try to harness the combined advantages of linear preconditioning schemes~\cite{Cai_2002, Baker_2005} and compressed representations of two-particle vertex functions~\cite{Kiese_dlr_2024, Shinaoka_2018}.

\emph{Acknowledgements.} The diagrammatic Monte Carlo data for our benchmark was provided by the authors of Ref.~\cite{Klett_2020}. The Flatiron Institute is a division of the Simons Foundation.  FK and KH have been supported by the Austrian Science Fund (FWF) projects  P36213, I5398,
SFB Q-M\&S (FWF project ID F86), and Research Unit QUAST by the Deutsche Foschungsgemeinschaft 
(DFG; project ID FOR5249) and FWF (project ID I 5868). IK has been supported by the European Research Council (ERC) under the European Union's Horizon 2020 research and innovation program (Grant No. 854843-FASTCORR). For the purpose of open access, the authors have applied a CC BY public copyright license to any Author Accepted Manuscript version arising from this submission. \newline

\appendix 

\section{SBE decomposition for the SU($2$)-symmetric single-band Hubbard model}
\label{app:sbe_chsps}

We here additionally state the SBE decomposition of the full four-point vertex $F$ in the `$\charge$', `$\spin$', `$\singlet$' notation
in the SU($2$)-symmetric case, cf. Eqs.~\eqref{eq:ch} to~\eqref{eq:si}, see Ref.~\cite{Krien_2019} for a detailed derivation.
Notice that the `\singlet' channel is not independent and can be expressed in terms of the `$\charge$' and `$\spin$'
channels using the relation $F^\singlet_{kk'q}=\frac{1}{2}(F^\charge_{kk',q-k-k'}-3F^\spin_{kk',q-k-k'})$,
\begin{widetext}
    \begin{subequations}
        \begin{align}
            F^\charge_{kk'q}=&\Lambda^{U,\charge}_{kk'q}+\Delta^{\charge}_{kk'q}
            -\frac{1}{2}\Delta^{\charge}_{k,k+q,k'-k}-\frac{3}{2}\Delta^{\spin}_{k,k+q,k'-k}
            +\frac{1}{2}\Delta^{\singlet}_{kk',q+k+k'}-2U,\label{eq:sbe_decomposition_ch}\\
            F^\spin_{kk'q}=&\Lambda^{U,\spin}_{kk'q}+\Delta^{\spin}_{kk'q}
            -\frac{1}{2}\Delta^{\charge}_{k,k+q,k'-k}+\frac{1}{2}\Delta^{\spin}_{k,k+q,k'-k}
            -\frac{1}{2}\Delta^{\singlet}_{kk',q+k+k'}+2U,\label{eq:sbe_decomposition_sp}\\
            F^\singlet_{kk'q}=&\Lambda^{U,\singlet}_{kk'q}+\Delta^{\singlet}_{kk'q}
            +\frac{1}{2}\Delta^{\charge}_{kk',q-k-k'}-\frac{3}{2}\Delta^{\spin}_{kk',q-k-k'}
            +\frac{1}{2}\Delta^{\charge}_{k,q-k',k'-k}-\frac{3}{2}\Delta^{\spin}_{k,q-k',k'-k}-4U.
            \label{eq:sbe_decomposition_si}
        \end{align}
    \end{subequations}
\end{widetext}
\section{Hedin equation with Fierz ambiguity}
\label{app:fierz}

Let us derive here the Hedin equation~\eqref{eq:hedin} including the Fierz parameter $0\leq r\leq1$, which allows to write the Hubbard interaction as
\begin{align}
    Un_\uparrow n_\downarrow=\frac{U}{2}\left[rnn+(r-1)mm\right]-\left(r-\frac{1}{2}\right)Un,
    \label{eq:fierz_identity}
\end{align}
where $n=n_\uparrow+n_\downarrow$ and $m=n_\uparrow-n_\downarrow$ are the operators for the density and magnetic moment.
In Eq.~\eqref{eq:fierz_identity} we consider only the easy-axis component, $s^z=\frac{m}{2}$, of the spin operator,
however, the following discussion holds analogously for a rotationally invariant splitting of the Hubbard interaction~\cite{Ayral_2016}.
The Hubbard Hamiltonian in the grand canonical ensemble can then be written in the form,
\begin{align}
    \mathcal{H}=H_0+\frac{1}{2}\sum_i \left[V^\charge n_in_i+V^\spin m_im_i\right]-(\mu+\mu_r)N.
    \label{eq:hubbard_fierz}
\end{align}
Here, $H_0$ describes the hopping, $N=\sum_i n_i$ is the total particle number operator.
$V^\charge=rU$ and $V^\spin=(r-1)U$ parametrize the splitting of the Hubbard interaction.
$\mu_r=\left(r-\frac{1}{2}\right)U$ takes the shift of the chemical potential due to the reformulation of the Hubbard interaction into account.
For the model in Eq.~\eqref{eq:hubbard_fierz} the equation of motion for the electronic Green's function takes the form~\cite{Krien_thesis}
\begin{align}
    &\left[\imath\nu-\varepsilon_k+\mu+\mu_r\right] - G_k^{-1}\notag\\
    =&V^\charge\langle n\rangle-\sum_{q,\alpha=\charge,\spin} G_{k+q}V^\alpha\gamma^{\text{red},\alpha}_{kq},
    \label{eq:eom_raw}
\end{align}
where $\gamma^{\text{red}}$ is an interaction-reducible three-leg vertex, related to the four-point vertex $F$ as,
\begin{align}
    \gamma^{\text{red},\alpha}_{kq}=1+\sum_{k'}G_{k'}G_{k'+q}F^\alpha_{k'kq}.
    \label{eq:gamma_red}
\end{align}
The shift $\mu_r$ could be absorbed into the noninteracting Green's function $G^0$, but here we move it to the self-energy $\Sigma$, which thus describes all changes induced by the middle ($V^\charge + V^\spin$) term on the right hand side of Eq.~(\ref{eq:hubbard_fierz}).
Then Eq.~\eqref{eq:eom_raw} reads,
\begin{align}
    \Sigma_k=-\mu_r+V^\charge\langle n\rangle-\sum_{q,\alpha=\charge,\spin} G_{k+q}V^\alpha\gamma^{\text{red},\alpha}_{kq},\label{eq:eom_raw2}
\end{align}
where we used $[G^0_k]^{-1}-G_k^{-1}=\Sigma_k$.
The interaction-reducible three-leg vertex is related to the Hedin vertex via
\begin{align}
    \gamma^{\text{red},\charge/\spin}_{kq}=\frac{W^{\charge/\spin}_q}{\pm U}\gamma^{\charge/\spin}_{kq}.
    \label{eq:gamma_red_hedin}
\end{align}
Using this together with the definition of $V^{\charge/\spin}$ we obtain the Hedin equation including the Fierz parameter,
\begin{align}
    \Sigma_k=&-\mu_r+rU\langle n\rangle\label{eq:hedin_fierz}\\
    -&\sum_{q}G_{k+q}\left[rW^\charge_q\gamma^{\charge}_{kq}-(r-1)W^\spin_q\gamma^{\spin}_{kq}\right]\notag.
\end{align}

\section{Resolution of the Fierz ambiguity through the crossing symmetry}
\label{app:fierz_sym}

We now show that a crossing-symmetric approximation for $\Lambda^U$ resolves the Fierz ambiguity.
Of course, to this end, we use in the following only identities which are satisfied by the self-consistent eMBEX solution.

Firstly, we recognize that the eMBEX self-energy is given by Eq.~\eqref{eq:hedin_fierz} with some $r$,
and that the Hedin vertex $\gamma$ is related to the four-point vertex $F$ via Eqs.~\eqref{eq:gamma_red_hedin} and~\eqref{eq:gamma_red}.
Hence, in eMBEX Eq.~\eqref{eq:hedin_fierz} is equivalent to 
\begin{align}
    \Sigma_k=&-\mu_r+\frac{U}{2}\langle n\rangle\label{eq:eom_fierz}\\
    -&U\sum_{qk'}G_{k+q}G_{k'}G_{k'+q}\left[rF^\charge_{k'kq}+(r-1)F^\spin_{k'kq}\right],\notag
\end{align}
where $F^{\charge/\spin}$ are given by the SBE decomposition in Eqs.~\eqref{eq:sbe_decomposition_ch} and~\eqref{eq:sbe_decomposition_sp}.
Notice that the one-body term in this expression is modified compared to
Eq.~\eqref{eq:hedin_fierz} due to the leading term $1$ in Eq.~\eqref{eq:gamma_red}.

It was recently shown~\cite{Yu_2024} that the SBE decomposition gives rise to a fluctuation decomposition of the self-energy,
$\Sigma=-\mu_r+\frac{U}{2}\langle n\rangle+\Sigma^{\text{2nd}}+\Sigma^\charge+\Sigma^\spin+\Sigma^\singlet+\Sigma^{\text{mb}}$,
where only the contribution of $\Lambda^U$ to Eq.~\eqref{eq:eom_fierz} depends explicitly on $r$:
\begin{align}
    \Sigma^{\text{mb}}_k=-U\sum_{qk'}G_{k+q}G_{k'}G_{k'+q}\left[r\Lambda^{U,\charge}_{k'kq}+(r-1)\Lambda^{U,\spin}_{k'kq}\right].\label{eq:eom_fierz_lambda}
\end{align}
The other terms, $\Sigma^{\text{2nd}}, \Sigma^\charge, \Sigma^\spin,$ and $\Sigma^\singlet$,
do not depend on $r$ and we ignore them in the following.
Notwithstanding the trivial term $\mu_r$,
the Fierz ambiguity is resolved if the $r$ dependence can be eliminated from Eq.~\eqref{eq:eom_fierz_lambda}.
This can be done using the crossing symmetry:
\begin{align}
    \Lambda^{U,\alpha}_{kk'q}=-\frac{1}{2}\left(\Lambda^{U,\charge}_{k,k+q,k'-k}+[3-4\delta_{\alpha,\spin}]\Lambda^{U,\spin}_{k,k+q,k'-k}\right).
    \label{eq:crossing_symmetry}
\end{align}
We consider the $r$-dependent part of the right-hand-side of Eq.~\eqref{eq:eom_fierz_lambda},
\begin{widetext}
    \begin{align}
        R_k
        =&-rU\sum_{qk'}G_{k+q}G_{k'}G_{k'+q}\left[\Lambda^{U,\charge}_{k'kq}+\Lambda^{U,\spin}_{k'kq}\right]\notag\\
        =&\frac{rU}{2}\sum_{qk'}G_{k+q}G_{k'}G_{k'+q}\left[\Lambda^{U,\charge}_{k,k+q,k'-k}+3\Lambda^{U,\spin}_{k,k+q,k'-k}\right]
        +\frac{rU}{2}\sum_{qk'}G_{k+q}G_{k'}G_{k'+q}\left[\Lambda^{U,\charge}_{k,k+q,k'-k}-\Lambda^{U,\spin}_{k,k+q,k'-k}\right]\notag\\
        =&rU\sum_{qk'}G_{k+q}G_{k'}G_{k'+q}\left[\Lambda^{U,\charge}_{k,k+q,k'-k}+\Lambda^{U,\spin}_{k,k+q,k'-k}\right].
    \end{align}
\end{widetext}
From the first to the second line we used the crossing symmetry~\eqref{eq:crossing_symmetry}.
After suitable shifts of the momentum-frequency summations we find $R_k=-R_k=0$ and the Fierz ambiguity is resolved.

We set $r=\frac{1}{2}$ in Eq.~\eqref{eq:hedin_fierz}, without loss of generality, leading to Eq.~\eqref{eq:hedin} in the main text,
which has the only advantage over other choices of $r$ that the slowly converging Matsubara sum
for the Hartree energy, ${U\langle n\rangle}/{2}$, is separated explicitly from the vertex correction $\gamma$.

More generally, the Fierz parameter $r$ drops out in any crossing-symmetric approximation for the full vertex $F$,
which eliminates $r$ already from Eq.~\eqref{eq:eom_fierz}.
As shown, in the eMBEX approach the crossing symmetry of $F$ is satisfied if it holds on the level of the fully $U$-irreducible vertex  $\Lambda^U$. One shows similarly that the Fierz ambiguity is resolved by the parquet approach~\cite{Bickers_2004,Toschi_2007,Kugler_2018_2,Krien_2021}, where the crossing symmetry of $F$ can be traced back to the corresponding fully $GG$-irreducible vertex function.

\section{Symmetries}
\label{app:symmetries}
\begin{figure*}[t]
    \centering
    \includegraphics[width = \linewidth]{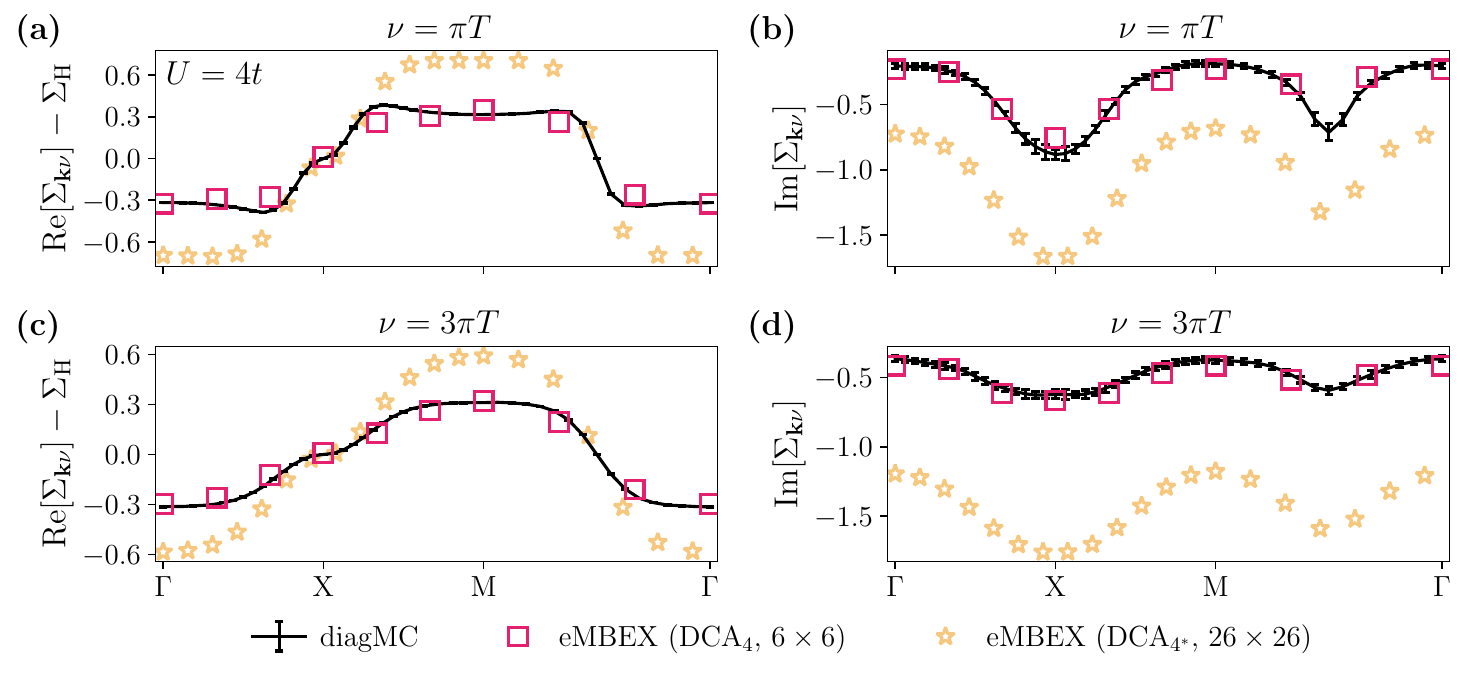}
    \caption{\textbf{Benchmark results for the half-filled Hubbard model at $U = 4t$ and $\beta t = 5$}. In this particular case, only eMBEX calculations using the irreducible vertex from $\DCAst$ could be converged on the large $26 \times 26$ lattice of the main text. By using a linear preconditioning technique~\cite{Cai_2002, Baker_2005} we converged $\DCAsq$ results on smaller lattices (here: $6 \times 6$).}
    \label{fig:sigma_benchmark_U4}
\end{figure*}
Besides the spin-rotation invariance, which was discussed in the main text, the Hubbard model possesses several additional symmetries which can be taken advantage of. Here, we show the effect of complex conjugation, time-reversal (TRS) and crossing symmetry on the self-energy, screened interactions, Hedin and irreducible vertices. A more detailed overview about vertex symmetries in the Hubbard model, including their derivation, can be found in Refs.~\cite{rohringer2013new, Rohringer_2018}.

Under complex conjugation the single-particle Green's function behaves as
\begin{align}
    \overline{G}_{\bm{k} \nu} = G_{\bm{k}, -\nu} \,,
\end{align}
which straightforwardly carries over to the self-energy, ${\overline{\Sigma}_{\bm{k} \nu} = \Sigma_{\bm{k}, -\nu}}$. Time-reversal symmetry, on the other hand, does not yield extra constraints on the one-particle level. For the screened interactions $W^{\alpha}$ the situation is similar, and we only exploit that
\begin{align}
    \overline{W}^{\alpha}_{\bm{q} \omega} = W^{\alpha}_{\bm{q}, -\omega} \,.
\end{align}
The Hedin vertices, in contrast, transform non-trivially under time-reversal. In particular, TRS permutes the fermionic four-vectors for any mixed momentum/frequency notation \cite{rohringer2013new}, which implies that only one of the two Hedin vertices per channel (see Eq.~\eqref{eq:delta}) needs to be considered and we, henceforth, limit ourself to $\gamma^{\alpha}$. Complex conjugation and crossing symmetry on the other hand imply that 
\begin{align} 
    \gamma^{\charge / \spin}_{k q} &= \overline{\gamma}^{\charge / \spin}_{-k, -q} \notag \\ 
    \gamma^{\charge / \spin}_{k q} &= \gamma^{\charge / \spin}_{q + k, -q} \,.
\end{align}
for the charge/spin channel, whereas the singlet channel transform as 
\begin{align} 
    \gamma^{\singlet}_{k q} &= \overline{\gamma}^{\singlet}_{-k, -q} \notag \\ 
    \gamma^{\singlet}_{k q} &= \gamma^{\singlet}_{q - k, q} \,,
\end{align}
For the irreducible vertex, complex conjugation and time-reversal symmetry yield the identities
\begin{align}
    \overline{\Lambda^{U, \alpha}}_{\bm{k} \bm{k}' \bm{q} \nu \nu' \omega } &= \Lambda^{U, \alpha}_{\bm{k}', \bm{k}, \bm{q}, -\nu', -\nu, -\omega} \\ 
    \Lambda^{U, \alpha}_{k k' q} &= \Lambda^{U, \alpha}_{k' k q} \,.
\end{align}
Since the crossing symmetry can be applied to the incoming or outgoing legs of $\Lambda^{U, \alpha}$ we find that 
\begin{align}
    \Lambda^{U, \alpha}_{k k' q} &= -\frac{1}{2} \left(\Lambda^{U, \charge}_{k, q + k, k' - k} + \left[3 - 4 \delta_{\alpha, \spin} \right] \Lambda^{U, \spin}_{k, q + k, k' - k} \right) 
    \label{eq:crossing1_chsp} \\ 
    &= -\frac{1}{2} \left(\Lambda^{U, \charge}_{q + k', k', k - k'} + \left[3 - 4 \delta_{\alpha, \spin} \right] \Lambda^{U, \spin}_{q + k', k', k - k'} \right) \,,
    \label{eq:crossing2_chsp}
\end{align}
for $\alpha \in \{ \charge, \spin \}$, whereas for the singlet channel 
\begin{align}
    \Lambda^{U, \singlet}_{k k' q} = \Lambda^{U, \singlet}_{q - k, k', q} = \Lambda^{U, \singlet}_{k, q - k', q} \,.
    \label{eq:crossing_s}
\end{align}
Note that all these functions are also invariant when transforming all of their respective momentum arguments according to the point group symmetries of the lattice under consideration. For the square lattice with point group $C_{4v}$, we implement $\tfrac{\pi}{2}$-rotation, ${(\bm{k}_x, \bm{k}_y) \to (-\bm{k}_y, \bm{k}_x)}$, and mirror symmetry about the $\tfrac{\pi}{4}$-diagonal, ${(\bm{k}_x, \bm{k}_y) \to (\bm{k}_y, \bm{k}_x)}$, since these operations are sufficient to generate the irreducible wedge of the Brillouin zone.

Since the result of our QMC simulation is exact up to numerical noise, all symmetries are fulfilled within the sampling error. Therefore, symmetries can be utilized to optimize the Monte Carlo data for the cluster and, of course, to reduce the numerical effort for converging the SBE equations on the lattice. For most of the relations discussed above, the implementation is straightforward. To enforce the crossing symmetry, however, we have to take into account that it can be used on the incoming legs, outgoing legs\FK{,} or both pairs of legs of the irreducible vertex. As is apparent from Eqs.~\eqref{eq:crossing1_chsp}, \eqref{eq:crossing2_chsp} and \eqref{eq:crossing_s}, each of these operations yields a separate relation between vertex components and as such, we have to factor in all of them in our symmetrization procedure. For the charge and spin channel, the symmetrized vertex reads:
\begin{align}
    \Lambda^{U, \alpha}_{k k' q} &= \frac{1}{4} \tilde{\Lambda}^{U, \alpha}_{k k' q}  + \frac{1}{4} \tilde{\Lambda}^{U, \alpha}_{q + k', q + k, -q} \notag \\ 
    &-\frac{1}{8} \left(\tilde{\Lambda}^{U, \charge}_{k, q + k, k' - k} + \left[3 - 4 \delta_{\alpha, \spin} \right] \tilde{\Lambda}^{U, \spin}_{k, q + k, k' - k} \right) \notag \\
    &-\frac{1}{8} \left(\tilde{\Lambda}^{U, \charge}_{q + k', k', k - k'} + \left[3 - 4 \delta_{\alpha, \spin} \right] \tilde{\Lambda}^{U, \spin}_{q + k', k', k - k'} \right) \,,
\end{align}
where $\tilde{\Lambda}^{U, \alpha}$ here corresponds to the \emph{raw} vertex from QMC. For the singlet channel we likewise have
\begin{align}
    \Lambda^{U, \singlet}_{k k' q} = \frac{1}{4} \tilde{\Lambda}^{U, \singlet}_{k k' q} + \frac{1}{4} \tilde{\Lambda}^{U, \singlet}_{q - k, q - k', q} + \frac{1}{4} \tilde{\Lambda}^{U, \singlet}_{q - k, k', q} + \frac{1}{4} \tilde{\Lambda}^{U, \singlet}_{k, q - k', q} \,.
\end{align}
Note that once the cluster vertex is symmetrized this way, our lattice parametrization from Sec.~\ref{sec:parametrization} likewise obeys the crossing symmetry, which makes our eMBEX implementation manifestly agnostic to the Fierz ambiguity (see the discussion in App.~\ref{app:fierz_sym}).

\section{eMBEX results for $U = 4t$}
\label{app:U4_benchmark}

Figure \ref{fig:sigma_benchmark_U4} shows the numerical results obtained for eMBEX calculations at $U = 4t$, where we found the SBE equations more difficult to converge. Without additional improvements to the algorithm outlined in Sec.~\ref{subsec:SBE_solver}, single-site and $\DCAsq$ calculations diverge after several iterations, even though the distance to the fixed point initially decreases. On the other hand, eMBEX with $\DCAst$ input \emph{can} be converged. The so-obtained solution has a rather large magnetic correlation length, and we argue that the convergence problems we experience for different $\Lambda^U$ inputs are therefore not due to a finite-size effect.

Indeed, by using a linear preconditioning technique~\cite{Cai_2002, Baker_2005}, the $\DCAsq$ calculation can be stabilized and we find that the converged solution on small grids (an example for a $6\times6$ momentum grid is shown in Fig.~\ref{fig:sigma_benchmark_U4}) is in good agreement with the benchmark data. Due to the increased computational cost of invoking the preconditioner as an intermediate step, larger grid sizes are currently not feasible and their exploration is left for future work. Note that the Hubbard model on the square lattice for $U/t=4$ and $\beta t = 5$ (at half-filling) is in a rather extreme regime, where a gap opens on the whole former Fermi surface and therefore, in retrospect, convergence problems with self-consistent schemes can be expected.

\begin{figure}[b]                                                     
    \begin{center}
        \includegraphics[width = 0.8\linewidth]{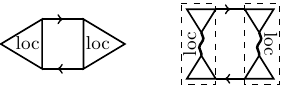}
    \end{center}
    \caption{\label{fig:polarization}\textbf{The effective exchange $J=\frac{t^2}{U}$} arises in DMFT from the diagram on the left, which includes the nonlocal four-point correction shown on the right.
    In this figure arrows denote the nonlocal DMFT Green's function $\tilde{G}$, see text.}
\end{figure}

\section{Magnetic correlations in single-site eMBEX}
\label{app:effective_exchange}

We discuss the diagrammatic content of eMBEX using a local $\Lambda^U$ as input,
$\Lambda^{U, \text{lat}}_{\bm{k}\bm{k'}\bm{q}\nu\nu'\omega} \approx {\Lambda}^{U,\text{loc}}_{\nu\nu'\omega}$.
First, we note that this approximation does not yield the same nonlocal two-particle correlations as DMFT,
which are obtained through inversion of the Bethe-Salpeter equation~\cite{Georges_1996,Hafermann_2014,Krien_2019_3}.
Instead, due to its crossing symmetry, eMBEX with $N_c=1$
recovers some nonlocal two-particle correlations which \emph{are not} included
in DMFT, while it also neglects others that \emph{are} included in DMFT.

In particular, the $N_c=1$ approximation does \emph{not}
recover the nearest-neighbor effective exchange energy scale $J=4t^2/U$ in the strong-coupling limit, which we can see as follows.
In DMFT, $J$ arises from the polarization diagram shown on the left of Fig.~\ref{fig:polarization}~\cite{Stepanov_2018,Krien_2019_3}.
Here, triangles denote the local spin-fermion coupling $\gamma^{\spin,\text{loc}}_{\nu\omega}$,
arrows the nonlocal DMFT Green's function $\tilde{G}_{\bm{k}\nu}=G^\text{DMFT}_{\bm{k}\nu}-g^\text{loc}_\nu$,
where $g^\text{loc}$ is the local Green's funcion of the DMFT's effective Anderson impurity model.

It is clear that diagrams included in ${\Lambda}^{U,\text{loc}}$ contain only local propagators $g^\text{loc}$.
However, the diagram on the left of Fig.~\ref{fig:polarization} contains four-point vertex corrections such as the one shown on the right,
where local single-boson exchange (see dashed boxes) is connected via nonlocal lines $\tilde{G}$;
wiggly lines denote the local screened interaction $w^\spin_\omega$.
This is a nonlocal two-boson exchange diagram, hence, it is not included in ${\Lambda}^{U,\text{loc}}$.
For this reason, while in the benchmarks we show this approximation for reference and comparison,
a cluster size of $N_c>1$ is mandatory in practical applications.

\bibliography{bib}
\end{document}